\documentclass[useAMS]{mn2e}
\usepackage{graphicx}
\usepackage{times,mathptm}
\newcommand{\ltsima}{$\; \buildrel < \over \sim \;$}
\newcommand{\simlt}{\lower.5ex\hbox{\ltsima}}
\newcommand{\gtsima}{$\; \buildrel > \over \sim \;$}
\newcommand{\simgt}{\lower.5ex\hbox{\gtsima}}

\newcommand{\lum}{\rm erg~s$^{-1}$}
\newcommand{\lx}{\rm $L_{2-10 keV}$}

\def\lesssim{\mathrel{\hbox{\rlap{\hbox{\lower4pt\hbox{$\sim$}}}\hbox{$<$}}}}
\def\gtrsim{\mathrel{\hbox{\rlap{\hbox{\lower4pt\hbox{$\sim$}}}\hbox{$>$}}}}

\def\arcsec{\hbox{$^{\prime\prime}$}}

\def\lxx{$\log~L_{\rm X}$}

\def\ab1450{$AB_{1450(1+z)}$}

\def\xray{\hbox{X-ray}}

\def\oiii{\hbox{[O\ {\sc iii}}]}

\def\lsun{\hbox{$L_\odot$}}
\def\loiii{$L_{\rm [O\ III]}$}
\newcommand\phn{\phantom{0}}% 

\def\edd_ratio{$\log\ L_{\rm bol}/L_{\rm Edd}$}

\def\l58{{$(\lambda L_{\lambda})_{\mbox{{\rm \scriptsize 5.8\micron}}}$}}
\def\lmir2{{$(\lambda L_{\lambda})_{\mbox{{\rm \scriptsize 12.3\micron}}}$}}
\def\s1{{S$_{\mbox{{\rm \scriptsize 3.6\micron}}}$}}
\def\irac2{{S$_{\mbox{{\rm \scriptsize 4.5\micron}}}$}}
\def\f3{{S$_{\mbox{{\rm \scriptsize 5.8\micron}}}$}}
\def\mic8{{S$_{\mbox{{\rm \scriptsize 8\micron}}}$}}
\def\asca{{\it ASCA\/}}
\def\chandra{{\it Chandra\/}}

\def\heao1{{\it HEAO-1\/}}

\def\iso{{\it ISO\/}}
\def\spitzer{{\it Spitzer\/}}

\def\rosat{{\it ROSAT\/}}

\def\sax{{\it BeppoSAX\/}}

\def\xmm{{XMM-{\it Newton\/}}}
\def\suzaku{{\it Suzaku\/}}

\def\swift{{\it Swift\/}}
\def\integral{{\it Integral\/}}
\def\aj{AJ}
\def\araa{ARA\&A}
\def\apj{ApJ}
\def\apjl{ApJ}
\def\apjs{ApJS}
\def\aap{A\&A}
\def\mnras{MNRAS}
\def\nat{Nature}
\def\procspie{Proc.~SPIE}
\title[THE CHANDRA QUEST FOR TYPE~2 QUASARS]
{Discovery of Compton-thick quasars in the Sloan Digital Sky Survey}
\author[C. Vignali et al.]
{
C. Vignali,$^{1,2}$\thanks{E-mail: cristian.vignali@unibo.it (CV); 
d.m.alexander@durham.ac.uk (DMA); roberto.gilli@oabo.inaf.it (RG); f.pozzi@unibo.it (FP).} 
D.~M. Alexander,$^{3}$\footnotemark[1] 
R. Gilli$^{2}$\footnotemark[1] and 
F. Pozzi$^{1}$\footnotemark[1] \\ \\ 
$^{1}$ Dipartimento di Astronomia, Universit\`a degli Studi di Bologna, 
Via Ranzani 1, 40127 Bologna, Italy \\
$^{2}$ INAF -- Osservatorio Astronomico di Bologna, Via Ranzani 1, 
40127 Bologna, Italy \\
$^{3}$ Department of Physics, Durham University, South Road, Durham DH1~3LE \\
}

\begin{document}
%\linenumbers*

\date{Accepted 2010 January 01. Received 2009 December 10; in original form 2009 June 26}

\pagerange{\pageref{firstpage}--\pageref{lastpage}} \pubyear{2010}

\maketitle

\label{firstpage}

\begin{abstract} 
We present new and archival \chandra\ snapshot ($\approx$~10~ks each)
observations of 15 optically identified (from the Sloan Digital Sky
Survey, SDSS) Type~2 quasars at \hbox{$z$=0.40--0.73}.
When combined with existing \xray\ data, this work provides complete 
\xray\ coverage for all 25 radio-quiet Type~2 quasars with 
log$L_{\rm [OIII]}>9.28$~\lsun\ from Zakamska et al. (2003). 
Two targets out of 15 were not detected by \chandra\ and most of the 
remaining sources are \xray\ weak, with nine having less than 10 counts 
in the 0.5--8~keV band. 
Low-to-moderate quality spectral analysis was limited to three sources, 
whose properties are consistent with the presence of column densities in 
the range $N_{\rm H}\approx$~10$^{22}$--10$^{23}$~cm$^{-2}$ in the 
source rest frame. 
If the \oiii\ luminosity is a reliable proxy for the intrinsic \xray\ 
luminosity, the current \xray\ data indicate that Compton-thick quasars 
may hide among $\approx$~65~per~cent of the SDSS Type~2 quasar population 
($L_{\rm \scriptsize X, meas}$/$L_{\rm \scriptsize X, [OIII]}$$<$0.01); 
however, since the Type~2 quasar sample is selected on \oiii\ luminosity, 
the estimated Compton-thick quasar fraction may be overestimated. 
Using archival \spitzer\ observations, we find that $\approx$~50~per~cent of 
SDSS Type~2 quasars appear to be obscured by Compton-thick material based 
on both the $L_{\rm \scriptsize X, meas}$/$L_{\rm \scriptsize X, mid-IR}$ 
(where mid-IR corresponds to rest-frame 12.3\micron) and 
$L_{\rm \scriptsize X, meas}$/$L_{\rm \scriptsize X, [OIII]}$ ratios. 
We use this information to provide an estimate of the Compton-thick quasar 
number density at $z\approx$~0.3--0.8, which we find is in broad agreement 
with the expectations from \xray\ background models.
\end{abstract}


\section{Introduction}
\label{intro}
Over the last decade, the quest for the identification of Type~2 
quasars\footnote{We defined Type~2 quasars as luminous (\lx$>10^{44}$~\lum) 
and obscured (column density $N_{\rm H}>10^{22}$~cm$^{-2}$) Active Galactic 
Nuclei - AGN - in X-rays, often characterized by high-ionization, 
narrow emission lines and the lack 
of broad emission lines in the rest-frame optical/ultraviolet 
spectra.} has received renewed interest, and many sources or promising 
candidates of this class have been detected, mainly through \xray\ surveys 
(e.g., Norman et al. 2002; Mainieri et al. 2002, 2007; Fiore et 
al. 2003; Gandhi et al. 2004, 2006; Mateos et al. 2005; Severgnini et 
al. 2006; Della Ceca et al. 2008; Krumpe et al. 2008). 
Predicted by unification schemes of AGN (e.g., Antonucci 1993), they 
are thought to play an important role in many synthesis models that 
seek to explain the origin of the \xray\ background (XRB; e.g., Gilli, 
Comastri \& Hasinger 2007).
On the other hand, the availability of deep radio and \spitzer\ 
observations up to 24\micron\ has allowed to define complementary 
methods to search for and characterize candidate Type~2 quasars (e.g., 
Martinez-Sansigre et al. 2005, 2008; Houck et al. 2005; Higdon et 
al. 2005, 2008; Weedman et al. 2006a,b,c; Dey et al. 2008; see also 
Donley et al. 2008 for a review), and examples of the most heavily 
obscured, Compton-thick AGN (with column density above 
$\approx$~10$^{24}$~cm$^{-2}$), have been possibly found (e.g., 
Alexander et al. 2005, 2008; Martinez-Sansigre et al. 2007; Daddi et al. 2007; 
Fiore et al. 2008, 2009, hereafter F09; Polletta et al. 2008; 
Lanzuisi et al. 2009).

An alternative approach to \xray\ and mid-infrared (mid-IR) surveys 
for the quest of Type~2 quasars consists in following-up, in the 
\xray\ band, promising Type~2 candidates selected through optical  
emission lines.  In this regard, a careful selection on the basis of 
the \oiii5007\AA\ luminosity allowed Zakamska et al. (2003; hereafter 
Z03) to define a sample of 291 Sloan Digital Sky Survey (SDSS; York et 
al. 2000) Type~2 quasar candidates in the redshift range 
$\approx$~0.3--0.8 (see also Zakamska et al. 2004, 2005, 2006 and 2008, 
and Liu et al. 2009 for observations and properties of these objects at other 
wavelengths). These sources are classified on the basis of 
high-excitation, narrow emission lines, without underlying broad 
components, and with line ratios characteristic of non-stellar 
ionizing radiation. Standard AGN photo-ionization models are able to 
successfully reproduced most of the emission-line ratio diagrams, at 
least for the most \oiii-luminous AGN in the Z03 sample 
(Villar-Mart{\'{\i}}n et al. 2008). 

In this context, the \xray\ band is crucial to confirm the AGN nature 
of these sources and assess the presence of obscuration, given all the 
uncertainties related to obscured AGN selection on the basis of the 
\oiii\ emission line (see $\S$\ref{number_density} for extended discussion). 
In Vignali, Alexander and Comastri (2004a, hereafter V04; see also 
Vignali, Alexander \& Comastri 2004b), we presented the basic \xray\ 
properties of a sub-sample of these Type~2 quasar candidates, mostly 
based on \rosat\ observations. In Vignali, Alexander \& Comastri 2006 
(hereafter V06), we placed significant constraints on the fraction of 
heavily obscured, possibly Compton-thick AGN among the population of 
SDSS Type~2 AGN, being of the order of $\approx$~50~per~cent (see also 
Ptak et al. 2006 for similar findings from an \xray\ perspective). 
We note that this result was possible by assuming the correlation 
between the \oiii\ and the \hbox{2--10~keV} flux found for Seyfert~2 
galaxies (Mulchaey et al. 1994; hereafter M94), which is characterized 
by a significant scatter.
%%%
Unlike the Type~2 quasar population found in moderately deep and 
ultra-deep \xray\ surveys, at the optical magnitude limit of the SDSS, 
Type~2 quasars are generally easy to study at both optical and \xray\ 
wavelengths, as confirmed by the results obtained over the last four 
years through a snapshot strategy with both \chandra\ and \xmm.

In this paper we extend the work shown by V04 and V06 by presenting 
\chandra\ observations for 12 additional SDSS Type~2 quasar candidates, 
thus providing a complete \xray\ coverage for all 25 radio-quiet 
Type~2 quasars with 
log$L_{\rm [OIII]}>9.28$~\lsun\ from Z03.  We have also included in 
our analyses three pointed \chandra\ observations retrieved from the 
public archive.  Coupled with previous \xray\ observations of AGN 
drawn from the original Z03 sample, the current targets bring the 
total number of sources with available \chandra\ and/or \xmm\ 
constraints to 31 (74~per~cent of \xray\ detections, but see also 
Lamastra et al. 2009 for recent updates).

The work by Z03 has been recently expanded and updated by 
Reyes et al. (2008; hereafter R08), where the \oiii5007\AA\ line luminosity 
is computed using both a Gaussian feature and a non-parametric line fitting 
method (see their Section~2.4). 
%besides the most up-to-date spectrophotometric calibration algorithm 
%from the SDSS 
More than 90~per~cent of the Type~2 quasar candidates reported by Z03 
were recovered from this more complex analysis. 
Since our project was originally thought to provide a reliable and complete 
\xray\ characterization of all the sources possibly classified as Type~2 
quasars from the Z03 sample - a goal pursued by specific runs of observations 
with \chandra\ (see V06 and this paper) - we will keep the original 
\oiii\ line luminosities reported by Z03 and discuss how the differences with 
respect to R08 impact on our conclusions on the number density of 
Compton-thick quasars in $\S$4.

%%%%%%%%%%%%%%%%%%%%%%%%%%%%%%%%%%%%%%%%%%%%%%%%%%%%%%%%%%%%
% V04: 17 sources, 3 X-ray detections (1 also XMM-Newton)
% V06: 16 sources, 10 X-ray detections (XMM-Newton included)
% V05: 15 sources, 13 X-ray detections
%%%%%%%%%%%%%%%%%%%%%%%%%%%%%%%%%%%%%%%%%%%%%%%%%%%%%%%%%%%%

The outline of the paper is as follows: in $\S$2 the SDSS Type~2 
quasar sample is presented, along with \chandra\ follow-up 
observations and \xray\ spectral results, while $\S$3 is 
focused on \spitzer\ observations and usage of the mid-IR luminosity 
as a proxy for the intrinsic \xray\ luminosity, 
hence to pick up heavily obscured quasars. 
The space density of SDSS Compton-thick quasars is shown and 
extensively discussed in $\S$4 in the light of current XRB model 
expectations and recent findings at higher redshifts. 
Finally, the main results are summarized in $\S$5. 

Hereafter we adopt the ``concordance'' (WMAP) cosmology 
($H_{0}$=70~km~s$^{-1}$~Mpc$^{-1}$, $\Omega_{\rm M}$=0.3, and 
$\Omega_{\Lambda}$=0.7; Spergel et al. 2003).

\section{Sample selection and Chandra observations}
\label{chandra_obs}
\begin{figure}
\includegraphics[angle=0,width=85mm]{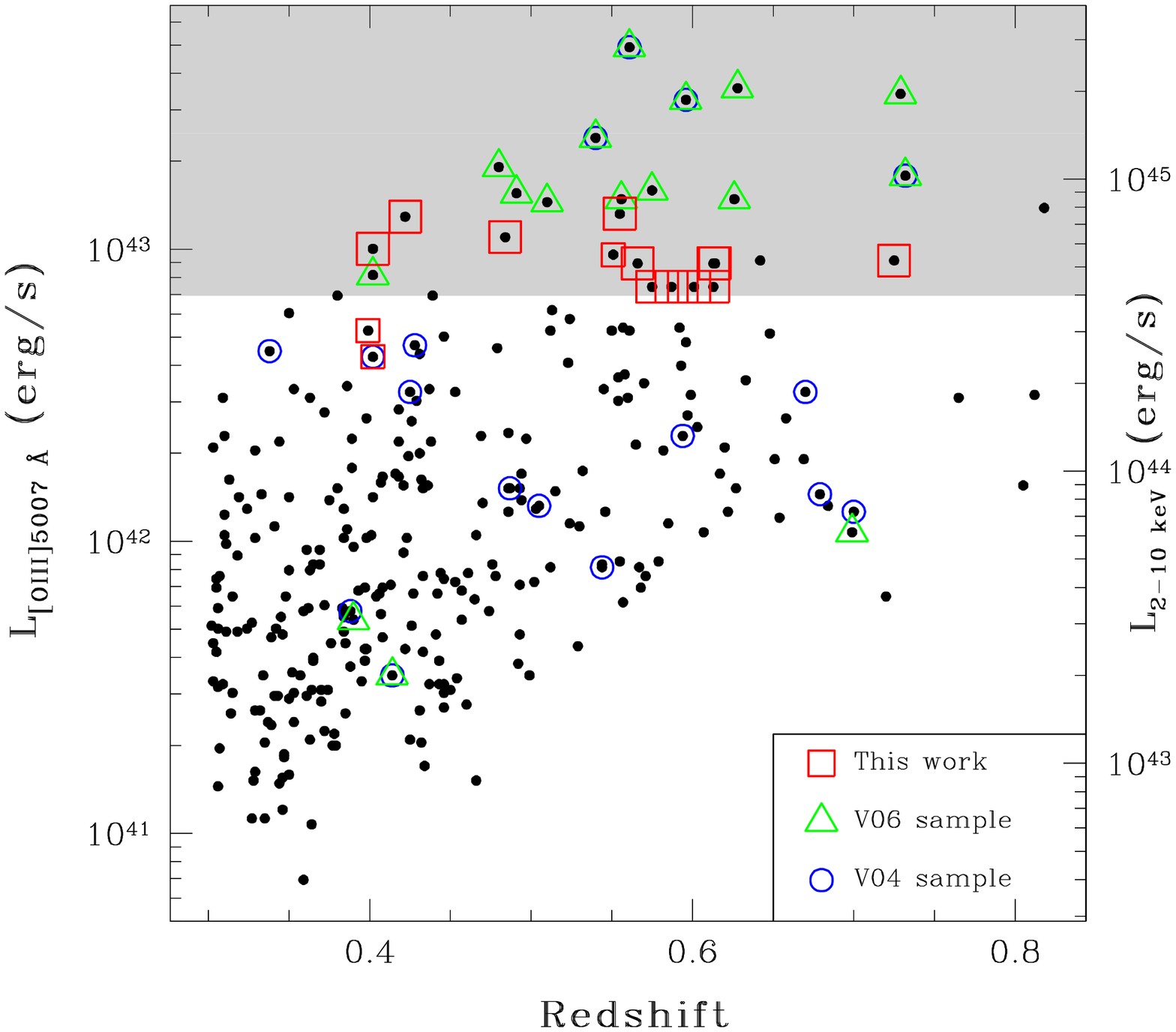}
\caption{
Logarithm of the measured \loiii\ luminosity (not corrected for 
extinction within the narrow-line region) vs. redshift for all of the 
sources in the Zakamska et al. (2003) catalog (small filled 
circles). At the right of the panel, the \hbox{2--10~keV} luminosity, 
estimated using the \oiii\ emission and the M94 correlation, are 
shown.  The key provides a description of the \xray\ observations; in 
particular, the sources presented in this paper are plotted as open 
squares (the three small open squares refer to the objects whose 
observations were retrieved from the \chandra\ archive). The grey 
region defines the locus of the AGN with \lxx $>44.6$~\lum, where the 
AGN still lie in the quasar luminosity range (\lxx$>44$~\lum) even 
assuming the 1$\sigma$ error around the mean of the original 
correlation between \oiii\ and hard X-rays.  We note that all of the 
objects in the grey region have been observed in X-rays except for two 
sources which are radio loud.}
\label{lxoiiiz}
\end{figure}

The 12 sources of the main sample presented in this work were 
specifically selected for \chandra\ Cycle~8 observations on the basis 
of their luminous \oiii\ emission.  Combined with previous sources 
targeted by \chandra\ (see V06 and Ptak et al. 2006), these 
observations provide a complete sampling of the luminous (i.e., in the 
quasar range) Type~2 AGN population revealed by the original study of 
Z03.  In particular, using the M94 correlation between \oiii\ and hard 
\xray\ flux, all the Z03 objects with predicted \hbox{2--10~keV} 
luminosity above $4\times10^{44}~$~\lum, i.e., in the quasar regime, 
have been observed by \chandra\ and \xmm\ (grey region in 
Fig.~\ref{lxoiiiz}).  The ``membership'' of all these sources to the 
``quasar locus'' (above 10$^{44}$~\lum) is valid even adopting the 
lowest \xray\ luminosities predicted from the M94 correlation assuming 
its 1$\sigma$ uncertainty, i.e., the sources included in the grey 
region of Fig.~\ref{lxoiiiz} have predicted hard \xray\ luminosities 
at least of 10$^{44}$~\lum. Only two sources in this ``quasar locus'' 
were not targeted, because they are radio loud (similarly to 
SDSS~081253.09$+$401859.9 of the archival sub-sample presented here), 
hence possibly not representative of the majority of the Type~2 quasar 
population, which is expected to be radio quiet.  The main sample of 
12 targets shown in Fig.~\ref{lxoiiiz} has been distinguished from the 
additional three sources retrieved from the \chandra\ archive on the 
basis of the size of the symbols (large and small open squares, respectively).
%Hereafter we refer to this sample as the ``main'' \chandra\ sample. 
%
%%%%%%%%%%%%%%%%%%%%%%%%%%%%%%%%%%%%%%%%%%%%%%%%%%%%%%%%%%%%%%%%%%%%%%%%%%
%%%	FIGURE 1: L[OIII] (erg/s) vs. Redshift for all of the Zakamska
%%%               et al. (2003) objects. Enlightened are the objects
%%%               with X-ray information in V04: 
%%%               open circles: X-ray upper limits
%%%               open triangles: X-ray detections
%%%               filled triangles: this work
%%%%%%%%%%%%%%%%%%
%%%%%%%%%%%%%%%%%%
%%%	END of FIG. 1
%%%%%%%%%%%%%%%%%%%%%%%%%%%%%%%%%%%%%%%%%%%%%%%%%%%%%%%%%%%%%%%%%%%%%%%%%%%

\subsection{Chandra data reduction and analysis}
\label{chandra_data}
The 12 SDSS Type~2 quasar candidates of the main sample were targeted 
by \chandra\ during Cycle~8 with the Advanced CCD Imaging Spectrometer 
(ACIS; Garmire et al. 2003) and the S3 CCD at the aimpoint; we used a 
snapshot strategy ($\approx$~10~ks exposures) and very faint mode for 
the event telemetry format.  The three archival observations, with 
similar exposures, were carried out during Cycle~7 in faint mode. 
Standard data reduction procedures were adopted using the \chandra\ 
Interactive Analysis of Observations ({\sc ciao}) Version~3.4 software. 
Source detection in the full \hbox{(0.5--8~keV)}, soft 
\hbox{(0.5--2~keV)} and hard \hbox{(2--8~keV)} energy bands was 
carried out with {\sc wavdetect} (Freeman et al. 2002) using a 
false-positive probability threshold of 10$^{-4}$, which is less 
conservative than that typically adopted in source detection in \xray\ 
survey fields, given the ``a priori'' knowledge of the source 
positions in our case.  Limited intervals of high background were 
present in two observations and removed through the generation of new 
good-time intervals.
The observation log is shown in Table~\ref{obs_log}; hereafter we 
will refer to the SDSS sources mostly using their abbreviated names. 
%%%%%%%%%%%%%%%%%%%%%%%%
%%%%%%%%%%%%%%%%%%%%%%%%%%%%%%%%%%%%%%%%%%%%%%%%%%%%%%%%%%%%%%%%%%%%%%%%%%
%%%	Table 1: Chandra observations of Type 2 QSOs: 
%%%              observation log (Cycle~8 targets)
%%%%%%%%%%%%%%%%
\begin{table*}
\centering
\begin{minipage}{140mm}
\caption{\chandra\ observation log for the SDSS Type~2 quasar candidates presented in this paper.}
%(Cycle~6 and 7).}
\label{obs_log}
%\scriptsize
\footnotesize
\begin{tabular}{rccccccc}
\hline
Src. ID \# & Object Name & $z$ & X-ray             & X-ray             & $\Delta_{\rm Opt-X}$ & 
Obs.~Date & Exp.~Time \\
           & SDSS~J      &     & ($\alpha_{2000}$) & ($\delta_{2000}$) & (arcsec)             & 
          & (ks)      \\
\hline
\multicolumn{8}{c}{\sf Sources observed as targets in Chandra AO8 -- Main Sample } \\ \\
  9 & 005621.72$+$003235.8 & 0.484 & 00 56 21.7 & $+$00 32 35.9 & 0.2      & 2008 Feb 08--09 &  9.91       \\
 16 & 012032.21$-$005502.0 & 0.601 & 01 20 32.2 & $-$00 55 02.0 & 0.2      & 2007 Feb 18     &  9.83$^{a}$ \\
 20 & 013416.34$+$001413.6 & 0.555 & 01 34 16.3 & $+$00 14 13.7 & 0.1      & 2007 Sep 10     &  9.91       \\ 
 29 & 014932.53$-$004803.7 & 0.566 & \dotfill   & \dotfill      & \dotfill & 2007 Aug 30     & 10.01       \\
 30 & 015716.92$-$005304.8 & 0.422 & 01 57 16.9 & $-$00 53 04.5 & 0.3      & 2007 Jun 18     &  9.92       \\
100 & 073745.88$+$402146.5 & 0.613 & 07 37 45.9 & $+$40 21 45.7 & 0.9      & 2007 Feb 03--04 &  9.33       \\ 
153 & 092152.45$+$515348.1 & 0.587 & \dotfill   & \dotfill      & \dotfill & 2007 Sep 27     & 10.04       \\ 
182 & 102746.03$+$003205.0 & 0.614 & 10 27 46.0 & $+$00 32 04.7 & 0.4      & 2007 Jan 13     &  9.91       \\
186 & 103951.49$+$643004.2 & 0.402 & 10 39 51.5 & $+$64 30 04.3 & 0.1      & 2007 Feb 04     &  9.52$^{a}$ \\
205 & 122845.74$+$005018.7 & 0.575 & 12 28 45.7 & $+$00 50 18.9 & 0.2      & 2007 Mar 12     &  9.42       \\
232 & 144642.29$+$011303.0 & 0.725 & 14 46 43.0 & $+$01 13 03.2 & 0.2      & 2007 Mar 22     & 10.04       \\
244 & 151711.47$+$033100.2 & 0.613 & 15 17 11.5 & $+$03 31 00.3 & 0.1      & 2007 Mar 28     &  9.91       \\
\hline 
\multicolumn{8}{c}{\sf Archival Sample} \\ \\ 
 18 & 012341.47$+$004435.9 & 0.399 & 01 23 41.5 & $+$00 44 35.9 & 0.2      & 2006 Feb 07--08 &  9.83       \\
117 & 081253.09$+$401859.9 & 0.551 & 08 12 53.1 & $+$40 19 00.5 & 0.6      & 2005 Dec 11     &  9.89       \\
152 & 092014.11$+$453157.3 & 0.402 & 09 20 14.1 & $+$45 31 56.9 & 0.4      & 2006 Mar 05     & 10.05       \\
\hline
\end{tabular}
\end{minipage}
%\hglue-0.3cm
\begin{minipage}[l]{140mm}
The source ID is taken from Table~1 of Z03. 
The optical positions of the quasars can be drawn from their 
SDSS name, while the \xray\ positions for the \xray\ detected sources have 
been obtained from either {\sc wavdetect} or the centroid of the source count 
distribution using the full-band image. 
We note that also the sources from the archival sample 
were targets of pointed \chandra\ observations (Cycle~7). 
%All of the exposure times were corrected for detector dead time. 
$^{a}$ Corrected for high-background periods. 
\end{minipage}
\end{table*}
%%%%%%%%%%%%%%%%
%%%	End of Table 1  
%%%%%%%%%%%%%%%%%%%%%%%%%%%%%%%%%%%%%%%%%%%%%%%%%%%%%%%%%%%%%%%

%%%%%%%%%%%%%%%%%%%%%%%%

In the total sample of 15 sources, only two targets were clearly not 
detected (SDSS~J0149$-$0048 and SDSS~J0921$+$5153), while all of the 
remaining sources were revealed by \chandra, although nine with less 
than 10 counts in the observed full band (see Table~\ref{chandra_counts}). 
These objects were carefully checked, 
following a procedure similar to that described in Vignali et 
al. (2001).  Monte-Carlo simulations indicate that all of the sources  
with less than 10 counts, including the \xray\ weak objects 
with 2--4~counts in the detection bands, are reliable detections 
(see last column of Table~\ref{chandra_counts}). 
From all these observations, coupled with those published by V06, we 
aim at deriving the basic \xray\ properties (e.g., \xray\ flux, hence 
luminosity, for all sources, and possibly photon index and absorption 
for the \xray\ brightest targets) of SDSS Type~2 quasars.

\subsection{X-ray spectral analysis}
\label{spectral_analysis}
Among the 15 targets analyzed in this paper, low-to-moderate quality 
\xray\ spectral analysis was possible only for three sources, 
having a number of full-band counts in the range \hbox{$\approx$~50--200}: 
SDSS~J1228$+$0050 (main sample), SDSS~J0123$+$0044 
and SDSS~J0812$+$4018 (archival sample; see Table~\ref{chandra_counts}). 

Spectral analysis was carried out with {\sc xspec} Version 11.3.2 
(Arnaud 1996) using unbinned data and the $C$-statistic (Cash 1979; 
Nousek and Shue 1989) for source SDSS~J1228$+$0050, while for the 
other two sources data were binned to at least 10 counts~per~bin. 
Errors are quoted at the 90~per~cent confidence level for one 
interesting parameter (\hbox{$\Delta$$C=2.71$}; Avni 1976; Cash 1979), 
unless stated otherwise.
Galactic absorption (from Dickey and Lockman 1990; see 
Table~\ref{xray_param}) was included in all the spectral fittings. 
%
%%%%%%%%%%%%%%%%%%%%%%%%
%%%%%%%%%%%%%%%%%%%%%%%%%%%%%%%%%%%%%%%%%%%%%%%%%%%%%%%%%%%%%%%%%%%%%%%%%%%
%%%	TABLE 2: X-ray photometry/wavdetect results [Chandra observations]
%%%%%%%%%%%%%%%%%%
\begin{table*}
\centering
\begin{minipage}{120mm}
\caption{X-ray counts, count rates, and results from Monte-Carlo simulations.}
\label{chandra_counts}
\footnotesize
\begin{tabular}{cccccc}
\hline
Source & \multicolumn{3}{c}{X-ray net counts}         & Count~rate   & Det. Sign.$^{a}$ \\
Name   & \cline{1-3} \\
SDSS~J & [0.5--2 keV] & [2--8 keV] & [0.5--8 keV] & [0.5--8 keV] &                  \\
\hline
0056$+$0032 &   2.9$^{+2.9}_{-1.6}$   & $<3.0$                 &  2.9$^{+2.9}_{-1.6}$     & 2.93 [1.31--5.85]  $\times10^{-4}$ & 3.6 \dotfill 2.9 \\
0120$-$0055 &   1.9$^{+2.6}_{-1.3}$   & 1.8$^{+2.6}_{-1.2}$    &  3.7$^{+3.1}_{-1.8}$     & 3.77 [1.93--6.92]  $\times10^{-4}$ & 4.0 3.3      4.8 \\
0134$+$0014 &   1.8$^{+2.6}_{-1.2}$   & $<4.8$                 &  2.7$^{+2.9}_{-1.5}$     & 2.73 [1.21--5.85]  $\times10^{-4}$ & 3.6 \dotfill 4.0 \\
0149$-$0048 &  $<4.8$                 & $<3.0$                 &  $<4.8$                  & $<$4.80 $\times10^{-4}$            &                 \\
0157$-$0053 &   5.9$^{+3.6}_{-2.4}$   & 16.7$^{+5.2}_{-4.0}$   & 22.7$^{+5.8}_{-4.7}$     & 2.29 [1.82--2.87]  $\times10^{-3}$ &                 \\ %2.7$\times10^{-10}$ 5.2$\times10^{-33}$ 4.6$\times10^{-44}$ \\
0737$+$4021 &   1.9$^{+2.6}_{-1.3}$   &   2.8$^{+2.9}_{-1.6}$  &    4.7$^{+3.3}_{-2.1}$   & 5.03 [3.00--8.58]  $\times10^{-4}$ & 3.7 4.1 5.5    \\
0921$+$5153 &  $<4.8$                 & $<3.0$                 &  $<4.8$                  & $<$4.78 $\times10^{-4}$            &                 \\
1027$+$0032 &   2.9$^{+2.9}_{-1.6}$   & $<3.0$                 &  2.7$^{+2.9}_{-1.5}$     & 2.72 [1.21--5.85]  $\times10^{-4}$ & 4.7 \dotfill 3.3 \\
1039$+$6430 &   4.0$^{+3.2}_{-1.9}$   & $<6.4$                 &  5.9$^{+3.6}_{-2.4}$     & 6.20 [3.68--9.98]  $\times10^{-4}$ & 5.7 \dotfill 5.7 \\
1228$+$0050 &   6.0$^{+3.2}_{-1.9}$   & 44.4$^{+7.7}_{-6.6}$   & 50.4$^{+8.2}_{-7.1}$     & 5.35 [4.60--6.22]  $\times10^{-3}$ &                  \\
1446$+$0113 &   5.9$^{+3.6}_{-2.4}$   & $<3.0$                 &  5.8$^{+3.6}_{-2.3}$     & 5.78 [3.49--9.37]  $\times10^{-4}$ & 6.4 \dotfill 5.5 \\
1517$+$0331 &   2.9$^{+2.9}_{-1.6}$   & $<3.0$                 &  2.7$^{+2.9}_{-1.5}$     & 2.72 [1.21--5.65]  $\times10^{-4}$ & 4.4 \dotfill 3.8 \\
\hline 
0123$+$0044 &  13.5$^{+4.8}_{-3.6}$   & 145.2$^{+13.1}_{-12.0}$& 161.2$^{+13.7}_{-12.7}$  & 1.64 [1.51--1.78]  $\times10^{-2}$ &                  \\
0812$+$4018 & 128.6$^{+12.4}_{-11.3}$ &  72.0$^{+9.5}_{-8.5}$  & 199.8$^{+15.2}_{-14.1}$  & 2.02 [1.88--2.17]  $\times10^{-2}$ &                  \\
0920$+$4531 &   6.8$^{+3.8}_{-2.5}$   & $<6.4$                 &   9.4$^{+4.2}_{-3.0}$    & 9.36 [6.37--13.54] $\times10^{-4}$ & 7.2 \dotfill 7.9 \\
\hline
\end{tabular}
\end{minipage}
\begin{minipage}{120mm}
The source names are abbreviated on the basis of the four RA and DEC digits. 
Errors on the \xray\ net counts (i.e., background-subtracted) 
were computed according to Gehrels (1986). 
The upper limits are at the 95\% confidence level and were computed 
according to Kraft, Burrows \& Nousek (1991). 
$^{a}$ Significance of the detection (in units of $\sigma$) in the soft band, hard band, and full band, respectively, 
estimated using the Monte-Carlo method described in Vignali et al. (2001) 
and applied only to sources with less than 10 counts. 
%For the X-ray faintest sources of the sample, the significance of the detection, 
%estimated using the Monte-Carlo method described in $\S$\ref{chandra_data}, 
%is reported (in units of $\sigma$) in the soft band, hard band, and full band, respectively. \\
%%%
\end{minipage}
\end{table*}
%%%%%%%%%%%%%%%%%%%%%
%%%	End of Table 2
%%%%%%%%%%%%%%%%%%%%%%%%%%%%%%%%%%%%%%%%%%%%%%%%%%%%%%%%%%%%%%%%%%%%%%%%%%%

%%%%%%%%%%%%%%%%%%%%%%%%
%

The limited energy band allowed by \chandra, coupled to the poor 
counting statistics, prevents us from an accurate estimate of the 
source spectral parameters. Having said that, we note that a fitting 
with a power-law model provides a flat photon index for all of the 
sources ($\Gamma=-0.31^{+0.40}_{-0.43}$, 
$\Gamma=-0.16^{+0.36}_{-0.37}$ and $\Gamma=1.22\pm{0.22}$ for 
SDSS~J1228$+$0050, SDSS~J0123$+$0044 and SDSS~J0812$+$4018, 
respectively).  In the light of the optical classification of these 
sources, the interpretation of the flatness of the 
\xray\ spectrum (if compared to that typically expected from Type~1 quasars; 
see, e.g., Piconcelli et al. 2005 and references therein) 
as due to absorption at  
the source provides a reliable representation of the \xray\ data. 
Indeed, the inclusion of absorption provides a more reliable fitting 
in term of resulting photon indices, which become consistent with 
those typically observed in Type~1 AGN (Page et 
al. 2005).\footnote{The resulting photon index is 
$\Gamma=1.9^{+1.6}_{-1.2}$ for SDSS~J1228$+$0050 and 
$\Gamma=2.6^{+0.5}_{-0.6}$ for SDSS~J0812$+$4018, while was fixed to 
$\Gamma=2.0$ for source SDSS~J0123$+$0044.} 
%%%%%%%%%%%%%%%%%%%%%%%
% 1228: 1.89+1.55/-1.22 
% 0123: 2.0 (frozen)
% 0812: 2.61+0.53/-0.55
%%%%%%%%%%%%%%%%%%%%%%%
The derived column densities are 
1.52$^{+1.30}_{-0.83}\times10^{23}$~cm$^{-2}$ (SDSS~J1228$+$0050), 
1.44$^{+0.49}_{-0.37}\times10^{23}$~cm$^{-2}$ (SDSS~J0123$+$0044) and 
2.14$^{+1.15}_{-0.81}\times10^{22}$~cm$^{-2}$ (SDSS~J0812$+$4018; see 
Table~\ref{xray_param}).  We note that the source with the lowest 
measured \xray\ absorption is SDSS~J0812$+$4018, which is also radio 
loud, with a radio loudness parameter\footnote{Radio-loudness 
parameter, defined as $R$ = $f_{\rm 5~GHz}/f_{\rm 
4400~\mbox{\scriptsize\AA}}$ (where 5~GHz and 4400~\AA\ are both rest 
frame; Kellermann et al. 1989).}  of $\approx$~4300. 
The interpretation of the \xray\ data for this source is complicated by 
its radio-loud nature; we note that in this case absorption could be 
ascribed to gas entrained by the unresolved radio jet (which may be 
responsible also for part of the \xray\ emission), although no firm 
conclusions can be drawn from the current \xray\ data. 
%
%%%%%%%%%%%%%%%%%%%%%%%%
%%%%%%%%%%%%%%%%%%%%%%%%%%%%%%%%%%%%%%%%%%%%%%%%%%%%%%%%%%%%%%%%%%%%%%%%%%
%%%	Table 3: Optical + X-ray properties
%%%%%%%%%%%%%%%%
\begin{table*}
\centering
\begin{minipage}{145mm}
\caption{Properties of the SDSS Type~2 quasars observed by \chandra.}
\label{xray_param}
%\scriptsize
\begin{tabular}{ccccccccc}
\hline
Object & $N_{{\rm H}_{\rm gal}}$ & $\log~L_{[OIII]}$ & $F_{2-10~keV}$
& $N_{{\rm H}_{\rm z}}$ & $L_{2-10~keV}$ & $L_{2-10~keV}$ (pr.) & \multicolumn{2}{c}{$S_{1.4~GHz}$} \\
\cline{8-9} \\
(1) & (2) & (3) & (4) & (5) & (6) & (7) & (8) & (9) \\
\hline                      %int    peak
0056$+$0032 & 2.79 & 9.45 & 1.0$\times10^{-15}$  & & 8.9$\times10^{41}$  & 1.5$\times10^{44}$--2.6$\times10^{45}$ &    8.60 &   8.33 \\
0120$-$0055 & 3.69 & 9.28 & 1.4$\times10^{-15}$  & & 2.1$\times10^{42}$  & 1.0$\times10^{44}$--1.8$\times10^{45}$ & \multicolumn{2}{c}{$<0.144$} \\
0134$+$0014 & 2.48 & 9.53 & 9.6$\times10^{-16}$  & & 1.2$\times10^{42}$  & 1.8$\times10^{44}$--3.2$\times10^{45}$ & \multicolumn{2}{c}{$<0.201$} \\
0149$-$0048 & 2.55 & 9.36 & $<1.7\times10^{-15}$ & & $<2.2\times10^{42}$ & 1.2$\times10^{44}$--2.1$\times10^{45}$ &    1.03 &   1.60 \\
0157$-$0053 & 2.58 & 9.52 & 8.1$\times10^{-15}$  & & 5.2$\times10^{42}$  & 1.8$\times10^{44}$--3.1$\times10^{45}$ & \multicolumn{2}{c}{$<0.135$} \\
0737$+$4021 & 6.18 & 9.28 & 1.9$\times10^{-15}$  & & 3.0$\times10^{42}$  & 1.0$\times10^{44}$--1.8$\times10^{45}$ & \multicolumn{2}{c}{$<0.134$} \\
0921$+$5153 & 1.27 & 9.28 & $<1.6\times10^{-15}$ & & $<2.3\times10^{42}$ & 1.0$\times10^{44}$--1.8$\times10^{45}$ &    2.37 &   2.49 \\
1027$+$0032 & 4.47 & 9.36 & 1.0$\times10^{-15}$  & & 1.6$\times10^{42}$  & 1.2$\times10^{44}$--2.1$\times10^{45}$ &    3.25 &   3.09 \\
1039$+$6430 & 1.18 & 9.41 & 2.1$\times10^{-15}$  & & 1.2$\times10^{42}$  & 1.4$\times10^{44}$--2.4$\times10^{45}$ & \multicolumn{2}{c}{$<0.5$} \\
1228$+$0050 & 1.88 & 9.28 & 1.8$\times10^{-14}$  & 1.52$^{+1.30}_{-0.83}\times10^{23}$  
                                                   & 3.5$\times10^{43}$  & 1.0$\times10^{44}$--1.8$\times10^{45}$ &    3.14 &   2.95 \\
1446$+$0113 & 3.55 & 9.37 & 2.1$\times10^{-15}$  & & 5.0$\times10^{42}$  & 1.3$\times10^{44}$--2.2$\times10^{45}$ &    5.75 &   5.75 \\
1517$+$0331 & 3.78 & 9.36 & 9.8$\times10^{-16}$  & & 1.5$\times10^{42}$  & 1.2$\times10^{44}$--2.1$\times10^{45}$ & \multicolumn{2}{c}{$<0.150$} \\
\hline  					    
0123$+$0044 & 3.24 & 9.13 & 5.8$\times10^{-14}$  & 1.44$^{+0.49}_{-0.37}\times10^{23}$ 
                                                   & 3.4$\times10^{44}$  & 7.3$\times10^{43}$--1.3$\times10^{45}$ &   11.83 &   11.34 \\
0812$+$4018 & 5.16 & 9.39 & 7.5$\times10^{-14}$  & 2.14$^{+1.15}_{-0.81}\times10^{22}$ 
                                                   & 1.7$\times10^{44}$  & 1.3$\times10^{44}$--2.3$\times10^{45}$ & 1067.88 & 1058.83 \\
0920$+$4531 & 1.51 & 9.04 & 3.2$\times10^{-15}$  & & 1.8$\times10^{42}$  & 5.9$\times10^{43}$--1.0$\times10^{45}$ & \multicolumn{2}{c}{$<0.5$} \\
\hline
\end{tabular}
\end{minipage}
\hglue1.5cm
\begin{minipage}{160mm}%{140mm}
(1) Abbreviated SDSS name; 
(2) Galactic column density, from Dickey \& Lockman (1990), in units of $10^{20}$~cm$^{-2}$; 
(3) log of the [OIII] line luminosity, in units of $L_{\odot}$ (from Z03); 
(4) Galactic absorption-corrected flux (or upper limit, in units of erg~cm$^{-2}$~s$^{-1}$) in the 2--10~keV band, extrapolated from the observed 
    \hbox{0.5--8~keV} count rate or upper limit assuming a power law with $\Gamma=2.0$ 
    (typical for AGN \xray\ emission). We note that assuming a pure cold reflection spectrum, 
    parameterized here as a power-law with $\Gamma=0$, would increase the {\it observed} 2--10~keV flux by a factor of $\approx$~8; 
    %or obtained directly from the \xray\ spectral fitting (when possible) with $\Gamma$ frozen to 2.0, for consistency with the other sources. 
    %The flux is in units of erg~cm$^{-2}$~s$^{-1}$; 
(5) intrinsic column density, obtained from the best-fit \xray\ spectral fitting, when possible, 
    %or using a power law with $\Gamma=2.0$, 
    in units of cm$^{-2}$; 
(6) 2--10~keV rest-frame luminosity, obtained through the observed flux (or upper limit), 
    corrected for the effect of intrinsic absorption (when the column density can be measured directly from the spectral fit), 
    in units of erg~s$^{-1}$; in case of a reflection spectrum, the 
    {\it rest-frame} 2--10~keV luminosity would increase by a factor of $\approx$~3. 
(7) 2--10~keV luminosity range predicted from the \oiii\ line vs. hard \xray\ flux correlation (M94), in units of erg~s$^{-1}$; 
    the interval is given by the $\pm{1}\sigma$ values around the mean M94 correlation; 
(8)--(9) integrated (8) and peak (9) radio flux density (or 1$\sigma$ upper limits) at 1.4~GHz, all from the FIRST 
         (Becker, White \& Helfand 1995) except for SDSS~J1039$+$6430, whose upper limit comes from the NVSS (Condon et al. 1998), in units of mJy. 
         We note that for SDSS~J1027$+$0032 the optical--radio association is dubious, since the 
         distance between the radio and optical position is large (1.0\arcsec). \\
All luminosities are computed using H$_{0}$=70~km~s$^{-1}$~Mpc$^{-1}$, 
$\Omega_{\rm M}$=0.3, and $\Omega_{\Lambda}$=0.7. 
\end{minipage}
\end{table*}
%%%%%%%%%%%%%%%%
%%%	End of Table 3
%%%%%%%%%%%%%%%%%%%%%%%%%%%%%%%%%%%%%%%%%%%%%%%%%%%%%%%%%%%%%%%

%%%%%%%%%%%%%%%%%%%%%%%%
%

For what concerns the remaining \xray\ detected sources, we were able 
to provide some constraints on the absorption only for 
SDSS~J0157$-$0053; its hardness ratio [defined as HR=(H$-$S)/(H$+$S), 
where H and S are the source counts in the hard and soft bands, 
respectively] of 0.48$\pm{0.32}$ is suggestive of a column density of 
\hbox{$\approx5\times10^{22}-10^{23}$~cm$^{-2}$}.

%%%%%%%%%%%%%%%%%%%%%%%%%%%%%%%%%%%%%%%%%%%%%%%%%%%%%%%%%%%%%%%%%%%%%%%%%%%
%%%	Figure 2: measured Lx (de-absorbed) vs. predicted Lx (according to
%%%		  the Mulchaey et al. 1994 correlation)
%%%%%%%%%%%%%%%%%%
\begin{figure}
\includegraphics[angle=0,width=85mm]{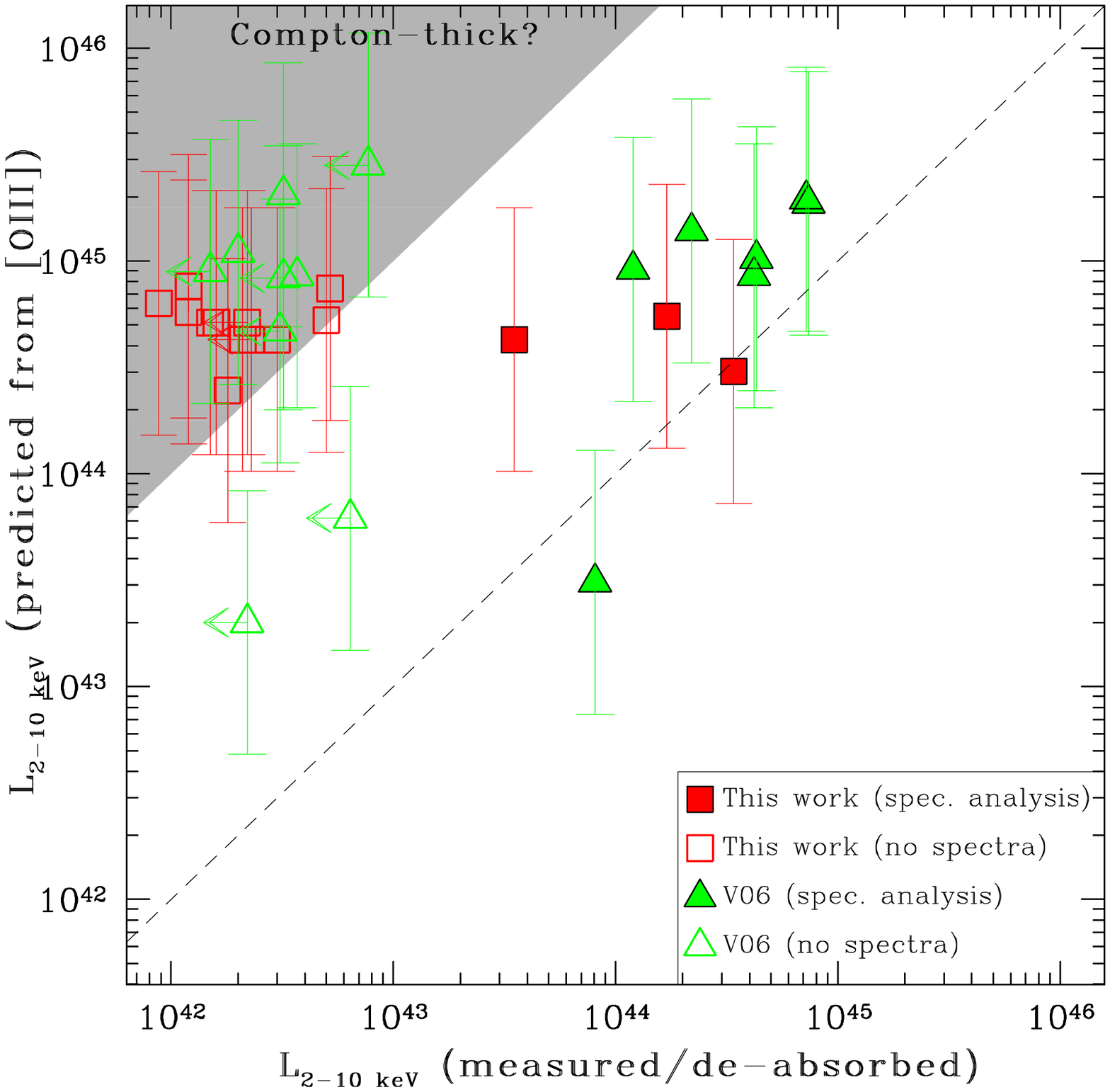}
\caption{
Comparison of the \hbox{2--10~keV} rest-frame luminosity computed from 
the available \xray\ data with that predicted assuming the M94 
correlation (i.e., derived from the \oiii\ luminosity). The dashed 
line shows the 1:1 ratio between the two \xray\ luminosities. 
All of the sources with either a \chandra\ or \xmm\ observation 
are shown. 
For the filled symbols (squares: data presented in this paper; 
triangles: data in V06), the \xray\ luminosity is de-absorbed on the 
basis of the column density derived from the best-fitting model (see 
Table~\ref{xray_param}); for all of the remaining sources, the \xray\ 
luminosity is derived from the \xray\ flux with no correction for the 
unknown column density.  Sources with \xray\ spectral results at the 
center of the plot are obscured, with typical measured column 
densities of a~few~$\times10^{22}$--5$\times10^{23}$~cm$^{-2}$, 
while the upper-left grey region shows the locus of extremely 
obscured, likely Compton-thick sources, where the observed luminosity 
is $\lesssim$1~per~cent of the predicted one.  Left-ward arrows 
indicate upper limits on the observed \xray\ luminosity.}
\label{complum}
\end{figure}
%%%%%%%%%%%%%%%%%%
%%%	END of FIG. 2
%%%%%%%%%%%%%%%%%%%%%%%%%%%%%%%%%%%%%%%%%%%%%%%%%%%%%%%%%%%%%%%%%%%%%%%%%%%

\subsection{X-ray results: discussion}
\label{discussion}
In this paper, as well as in V06, we have shown that the population of 
SDSS Type~2 quasar candidates investigated by \chandra\ is 
characterized by low \xray\ emission, in comparison with that expected 
from the \oiii5007\AA\ emission-line intensity and the M94 
correlation.  
In the following discussion, we consider all of the sources with 
either a \chandra\ or \xmm\ observation. 
We note that for six sources for which we refer to 
\chandra\ data, \xmm\ observations are also present in the archive; 
in particular, four of these six objects were observed as targets by \xmm. 
Of these six sources, we have only one detection; for the other sources, 
the integration time is typically lower or comparable to that of \chandra\ 
observations, once the intervals of high particle background (flares) 
are removed. 
This means that the constraints achieved using \chandra\ data for these 
six sources remain the best available to date. 
About the source detected by \xmm, it was already detected by 
\chandra\ (src.~\#15; V06) and was bright enough for an \xray\ spectral 
extraction. 
However, we note that the presence of 
diffuse \xray\ emission in the \xmm\ observation within the aperture adopted 
for \xray\ spectral extraction may provide some contamination to the AGN 
properties, as shown by the source flux which is $\approx$~70~per~cent 
higher than in the \chandra\ observation. 
For this reason, we prefer to use the 
\chandra\ constraints (as reported in V06) also for this source.

If we correct the \xray\ luminosity accounting for the 
measured absorption (when a column density estimate is feasible 
through spectral analysis, see the filled squares and triangles in 
Fig.~\ref{complum}; the measured column densities are in the range 
$\approx1.5\times10^{22}-4.3\times10^{23}$~cm$^{-2}$, see V06 and 
$\S$\ref{spectral_analysis}), we generally find a good agreement 
between the expected and the observed 2--10~keV luminosity.  On the 
other hand, there is a significant fraction of Type~2 AGN candidates 
($\approx$~65~per~cent) whose observed luminosity is less than 
1~per~cent of the predicted one; for these sources, heavy obscuration 
towards the nuclear source appears a reasonable explanation for their 
weak \xray\ emission, supported by the optical classification of Z03 
(see also Ptak et al. 2006 for similar conclusions, 
and the recent results of LaMassa et al. 2009 using a lower redshift 
sample of SDSS Seyfert~2 galaxies).  Hereafter, we 
will assume the sources with measured/predicted \xray\ luminosity less than 
0.01 as probably Compton thick. This is a conservative threshold 
since there are known Compton-thick AGN in the local Universe with 
more than 1~per~cent of observed over intrinsic \xray\ emission (e.g., 
NGC~6240 and the Circinus galaxy; Bassani et al. 1999; Matt et al. 2000; 
see also Maiolino et al. 1998, where this fraction is $\approx$~2~per~cent 
for a sample of Compton-thick Seyfert~2 galaxies).
%%%%%%%%%%%%%%%%%%%%%%%%%%%%%%%%%%%%%%%%%%%%%%%%%%%%%%%%%%%%%%%%%%%%%%%%%
% Circinus: Matt+99: 3.4e41-1.7e42 cgs [2-10 keV, intrinsic], BeppoSAX
%           Yang+09: 1.1e42 [2-10 keV] vs. 2.4e42 [2-100 keV], Suzaku
% NGC 6240: Vignati+99: 3.6e44 cgs [2-10 keV, intrinsic], BeppoSAX, <1%?
%%%%%%%%%%%%%%%%%%%%%%%%%%%%%%%%%%%%%%%%%%%%%%%%%%%%%%%%%%%%%%%%%%%%%%%%%

Similarly to V06, we did not correct the \oiii\ flux to account for 
the absorption due the narrow-line region (NLR) itself (for details, 
see Maiolino et al. 1998; Bassani et al. 1999; Panessa et al. 2006, 
hereafter P06; see also Lamastra et al. 2009), because the SDSS 
spectral coverage does not typically cover the rest-frame wavelengths 
necessary to measure the emission-line fluxes of H$\beta$ and 
H$\alpha$, hence to use the H$\alpha$/H$\beta$ ratio as extinction 
measurement.  Using the ratio of higher order Balmer lines (e.g., 
H$\beta$/H$\gamma$, H$\gamma$/H$\delta$) to estimate extinction 
represents a difficult task, due to their weakness over often noisy 
spectra.  This approach implies that the predicted \xray\ luminosities 
would be even higher for at least some sources, thus generally 
requiring a larger column density to account for the observed luminosities.

Furthermore, we have assumed the correlation between the \oiii\ 
emission-line intensity and the hard \xray\ emission from M94, which 
is based on a large sample ($>100$) of local Seyfert galaxies.  
As reported in Table~\ref{xray_param}, our objects have \xray\ 
luminosities well below those predicted by the M94 correlation; 
13 out of the 15 targets have \xray\ measured/predicted luminosity ratios 
($L_{\rm \scriptsize X, meas}$/$L_{\rm \scriptsize X, [OIII]}$)
in the range $\approx0.001-0.081$ 
(where for the predicted \xray\ luminosity, the ``central'' value from the 
M94 correlation has been assumed). 

A few caveats must be considered before comparing different \oiii\ 
vs. hard \xray\ correlations.  If a pure cold reflection spectrum, roughly 
parameterized here as a flat ($\Gamma=0$) power-law slope, were 
present in the observed 0.5--8~keV band covered by \chandra, the 
2--10~keV fluxes reported in Table~\ref{xray_param} would increase by 
a factor $\approx$~8, and the rest-frame 2--10~keV luminosities by a 
factor $\approx$~3.  Still a significant ($\approx$~40--50~per~cent) 
fraction of SDSS Type~2 quasars would populate the Compton-thick locus 
in Fig.~\ref{complum}.  We note, however, that studies of local 
reflection-dominated AGN indicate that a scattering component is 
possibly present in such cases (e.g., Comastri 2004), mostly 
``contaminating'' the soft \xray\ band; this would produce softer 
spectra and, as a consequence, any column density estimated through 
low-to-moderate quality \xray\ spectral information would be under-estimated. 

Since M94 work, many authors have searched for the presence of a 
correlation between the \oiii\ emission-line intensity and the hard 
X-rays. H05 clearly show the differences in deriving such correlation 
from an optically selected sample.  In particular, they found that 
$\log$(\lx/\loiii) is 1.59 (dispersion $\sigma$=0.48) for optically 
(\oiii) selected Type~1 AGN, while $\log$(\lx/\loiii)=0.57 (dispersion 
$\sigma$=1.06) for optically selected Type~2 AGN.  We remind that the 
best-fit ratio assumed throughout this paper is 
\lx/\loiii$\approx$~1.8 ($\sigma\approx$~0.6), similar for Type~1 and 
Type~2 AGN (M94); this value is not significantly different from that 
of Type~1 AGN in H05.  Under the basic assumption (unified schemes) 
that the intrinsic \lx/\loiii\ should be the same for Type~1 and 
Type~2 AGN, the only difference being the obscuration/extinction of 
the nuclear emission, to report our predicted \xray\ luminosities 
(reported in Table~\ref{xray_param} and shown in Fig.~\ref{complum}) 
into those of H05 we have to divide them by $\approx$~1.5.  Given the 
significant dispersion of H05 correlation, a large fraction of sources 
($\approx$~50~per~cent) would still be in the locus of heavily 
obscured, possibly Compton-thick quasars. 

Finally, we note that the correlation found by P06 produces even 
larger hard \xray\ luminosities than those reported in this paper (see 
also $\S$\ref{number_density}). 
For consistency with the assumptions of V04 and V06 which, combined 
with the present work, provide a complete \xray\ coverage of all 
radio-quiet Type~2 quasars with log$L_{\rm [OIII]}>9.28$~\lsun\ 
selected from Z03, we have decided to adopt the M94 correlation, 
keeping in minds all the caveats concerning both the selection 
criteria of the different samples and the significant dispersion in 
all of the correlations presented above; for further discussion, refer 
to $\S$\ref{number_density}.  Support to the reliability of the 
\oiii-X-ray correlation is also provided by \spitzer\ data (see 
$\S$\ref{calib_mir}).
%---------------------------------------------------------------
%M94: Log (Lhx/Loiii)=1.82, similar for Type~1 and Type~2 AGN \\
%H05: Log (Lhx/Loiii)=1.59, $\sigma$=0.48 for Type~1 AGN and \\
%     Log (Lhx/Loiii)=0.57, $\sigma$=1.06 for Type~2 AGN. \\
%---------------------------------------------------------------

\section{Near and mid-IR observations of SDSS Type~2 quasars}
\label{calib_mir}

\subsection{Spitzer data reduction and analysis}
\label{spitzer_data}
In the following, we estimate the nuclear power of the AGN by means of 
the rest-frame 5.8\micron\ and 12.3\micron\ luminosities, providing a 
robust support to the results from \xray\ analysis.  To this purpose, 
we retrieved from the \spitzer\ archive both IRAC and MIPS photometric 
observations for 20 SDSS Type~2 quasars in common with those reported 
in this work and in V04a and V06. Most of these sources were targeted by 
\spitzer\ to get insights, for the first time, on the near-IR and mid-IR 
properties of a well defined sample of optically selected Type~2 AGN. 

In particular, by these infrared data, we 
intend to provide an adequate complement to the \xray\ investigation 
presented in the previous sections.  The \xray\ properties of these 
sources have been published mostly in V06 and Ptak et al. (2006); for 
seven of these sources, \xray\ data are presented here.  While IRAC 
covers the near-infrared (near-IR) and mid-IR bands with four channels 
(3.6, 4.5, 5.8 and 8\micron), MIPS allows for investigations at longer 
wavelengths (24, 70 and 160\micron), although in this work only 
24\micron\ data have been used. 

For both IRAC and MIPS, we used the final post-calibrated data (PBCD) 
produced by the \spitzer\ Science Center (SSC) pipeline (Version 
14.0.0 and 16.1.0, respectively).  The flux densities in the IRAC and 
MIPS bands were measured on the signal maps using aperture photometry. 
The chosen aperture radius for the IRAC bands was 2.45\arcsec, and the 
adopted factors for the aperture corrections are 1.21, 1.23, 1.38 and 
1.58 (following the IRAC Data Handbook) at 3.6\micron, 4.5\micron, 
5.8\micron, and 8\micron, respectively.  The aperture radius for the 
24\micron\ MIPS band was 7\arcsec, and the resulting aperture flux 
density was corrected adopting a factor of 1.61 (see the MIPS Data Handbook).

All of the quasars have been detected by \spitzer, spanning a flux 
density range at 8\micron\ (24\micron) of $\approx$~0.11~mJy 
($\approx$~1.46~mJy; source ID=290, the faintest object in the current 
sample) to $\approx$~12.8~mJy ($\approx$~56.4~mJy; source ID=197, the 
brightest source).  Only four sources show indications of limited 
contamination by a nearby companion; however, in all of these cases 
the nearby source shows a decreasing flux density at longer 
wavelengths, suggesting a stellar or galaxy origin for its 
near-IR/mid-IR emission.  The rest-frame 5.8\micron\ and 12.3\micron\ 
luminosities (Table~\ref{spitzer}) were derived via simple 
interpolation between nearby bands in the source rest frame; this 
method avoids that the results are dependent upon the assumption of 
particular AGN (torus) templates to reproduce the observed data 
points.  Furthermore, we checked our results against the 14.5\micron\ 
luminosities obtained using IRS spectra and reported by Zakamska et 
al. (2008) for eight sources in common, and found a relatively good 
agreement.  The combined statistical and systematic errors associated 
to these mid-IR luminosities are \hbox{$\approx$~10--15~per~cent}. 
In Table~\ref{spitzer}, also basic \xray\ information and corresponding 
reference paper are presented. 
%%%%%%%%%%%%%%%%%%%%%%%%
%%%%%%%%%%%%%%%%%%%%%%%%%%%%%%%%%%%%%%%%%%%%%%%%%%%%%%%%%%%%%%%%%%%%%%%%%%
%%%	Table 4: Infrared + X-ray properties
%%%%%%%%%%%%%%%%
\begin{table*}
\centering
\begin{minipage}{125mm}
\caption{Properties of the SDSS Type~2 quasars observed by \spitzer\ IRAC and MIPS.}
\label{spitzer}
\begin{tabular}{ccccccc}
\hline
ID  & Src. Name & $z$ &  log \l58 & log \lmir2 & X-ray info & log~L$_{\rm X}$ \\
    &           &     & (\lsun)   & (\lsun)    &            &                 \\  
(1) & (2)       & (3) &  (4)      & (5)        & (6)        & (7)             \\
\hline
  9 & 005621.72$+$003235.8 & 0.484 & 11.3 & 11.6 &  D/C/V09	& 41.95 \\
 18 & 012341.47$+$004435.9 & 0.399 & 11.2 & 11.2 &  D+/C/V09    & 44.53 \\
 20 & 013416.34$+$001413.6 & 0.555 & 11.2 & 11.5 &  D/C/V09	& 42.08 \\
 30 & 015716.92$-$005304.8 & 0.422 & 10.7 & 10.9 &  D/C/V09	& 42.72 \\
 34 & 021047.01$-$100152.9 & 0.540 & 11.2 & 11.3 &  D/X/V06	& 44.34 \\
 83 & 031950.54$-$005850.6 & 0.626 & 11.0 & 11.2 &  D/C/V06	& 42.57 \\
113 & 080154.24$+$441234.0 & 0.556 & 11.2 & 11.4 &  D+/C/V06    & 44.62 \\
117 & 081253.09$+$401859.9 & 0.551 & 11.4 & 11.4 &  D+/C/V09    & 44.23 \\
119 & 081507.42$+$430427.2 & 0.510 & 11.8 & 11.8 &  ND/C/V06    & $<42.51$ \\
130 & 084234.94$+$362503.1 & 0.561 & 10.9 & 11.4 &  ND/C/V06    & $<42.89$ \\
152 & 092014.11$+$453157.3 & 0.402 & 11.3 & 11.5 &  D/C/V09	& 42.26 \\
186 & 103951.49$+$643004.2 & 0.402 & 11.6 & 11.7 &  D/C/V09	& 42.08 \\
196 & 115314.36$+$032658.6 & 0.575 & 11.2 & 11.3 &  D+/C/V06    & 44.08 \\
197 & 115718.35$+$600345.6 & 0.491 & 12.1 & 12.2 &  ND/C/V06    & $<42.18$ \\
204 & 122656.48$+$013124.3 & 0.732 & 11.6 & 11.6 &  D+/X/V06    & 44.63 \\
207 & 123215.81$+$020610.0 & 0.480 & 11.1 & 11.3 &  D/C/V06	& 42.30 \\
239 & 150117.96$+$545518.3 & 0.338 & 10.7 & 10.9 &  D/R/V04	& 44.24 \\
256 & 164131.73$+$385840.9 & 0.596 & 11.6 & 11.8 &  D+/X/V06    & 44.87 \\
289 & 235818.87$-$000919.5 & 0.402 & 10.6 & 10.9 &  ND/C/V06    & $<42.49$ \\
290 & 235831.16$-$002226.5 & 0.628 & 10.3 & 10.8 &  D/C/V06	& 42.51 \\
\hline
\end{tabular}
\end{minipage}
\hglue-0.2cm
\begin{minipage}{120mm}
(1) Source ID (from Table~1 of Z03); 
(2) SDSS source name; 
(3) source redshift; 
%(4)--(7) flux density in the IRAC bands, in units of \mjy; 
%(8) flux density in the MIPS 24~\micron\ band, in units of \mjy; 
(4) log of the luminosity at 5.8~\micron\ (rest frame), in units of \lsun;
(5) log of the luminosity at 12.3~\micron\ (rest frame), in units of \lsun; 
(6) available \xray\ information: the first term indicates whether the source has been detected in X-rays 
(D: detected; D+: detected and \xray\ spectrum available; ND: undetected); 
the second term indicates the \xray\ satellite whose data have been used to obtain information 
(C: \chandra; X: \xmm; R: \rosat/RASS). 
The last term indicates the reference paper (where V09 means the current work); 
(7) logarithm of the rest-frame 2--10~keV luminosity (\lum) or upper limit when the source is undetected. 
When an \xray\ spectrum is available, the \xray\ luminosity has been corrected to account for the 
measured absorption. 
\end{minipage}
\end{table*}
%%%%%%%%%%%%%%%%
%%%    End of Table 4
%%%%%%%%%%%%%%%%%%%%%%%%%%%%%%%%%%%%%%%%%%%%%%%%%%%%%%%%%%%%%%%

%%%%%%%%%%%%%%%%%%%%%%%%

%%%%%%%%%%%%%%%%%%%%%%%%%%%%%%%%%%%%%%%%%%%%%%%%%%%%%%%%%%%%%%%%%%%%%%%%%%%%%%%%%%%%%
%%%	FIGURE 3: Measured Lx vs. expected Lx from the 5.8 and 12.3 micron luminosity 
%%%               [Fiore+08 and Gandhi et al. relations]
%%%%%%%%%%%%%%%%
\begin{figure*}
\includegraphics[angle=0,width=0.48\textwidth]{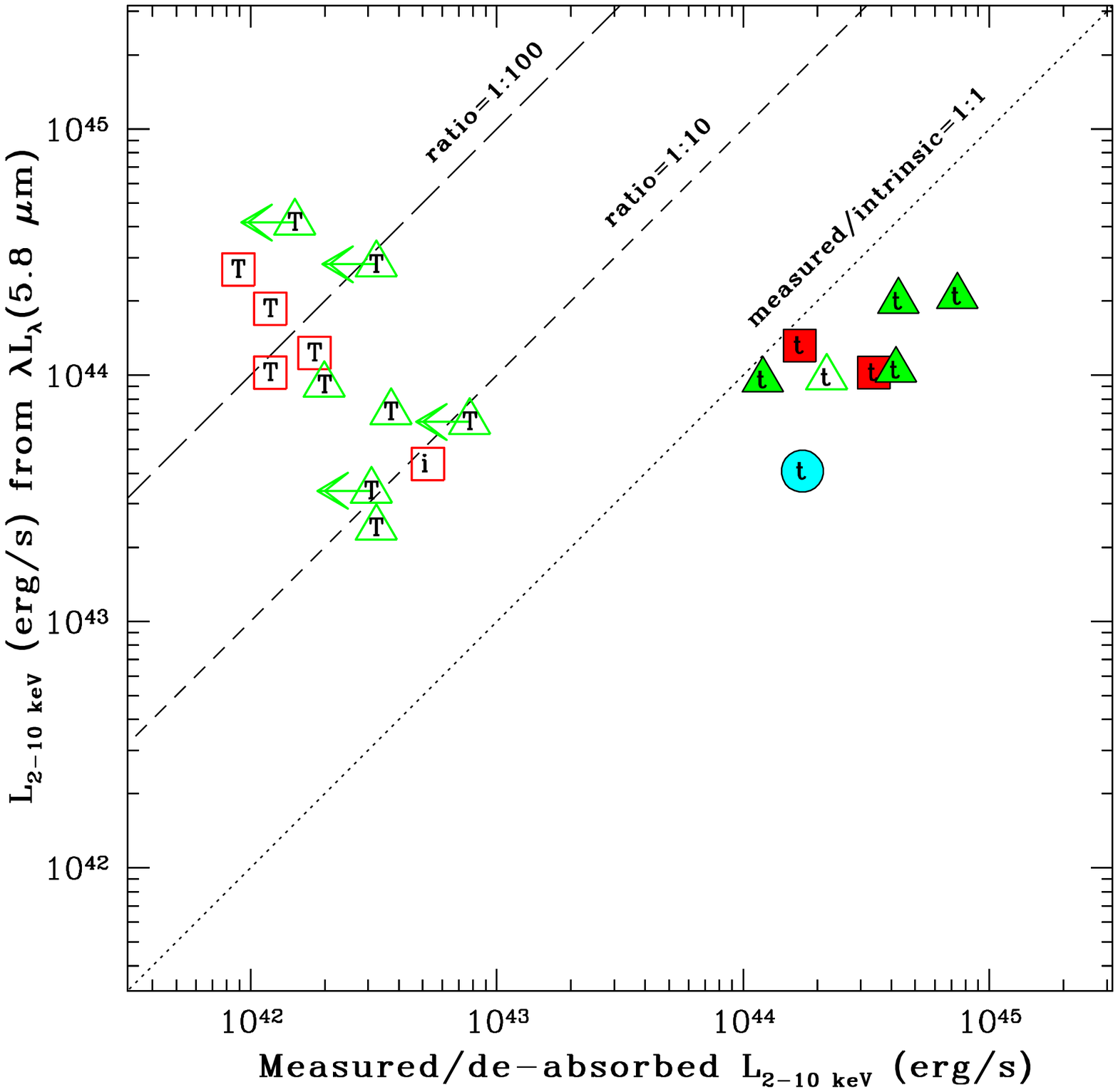}
\hfill
\includegraphics[angle=0,width=0.48\textwidth]{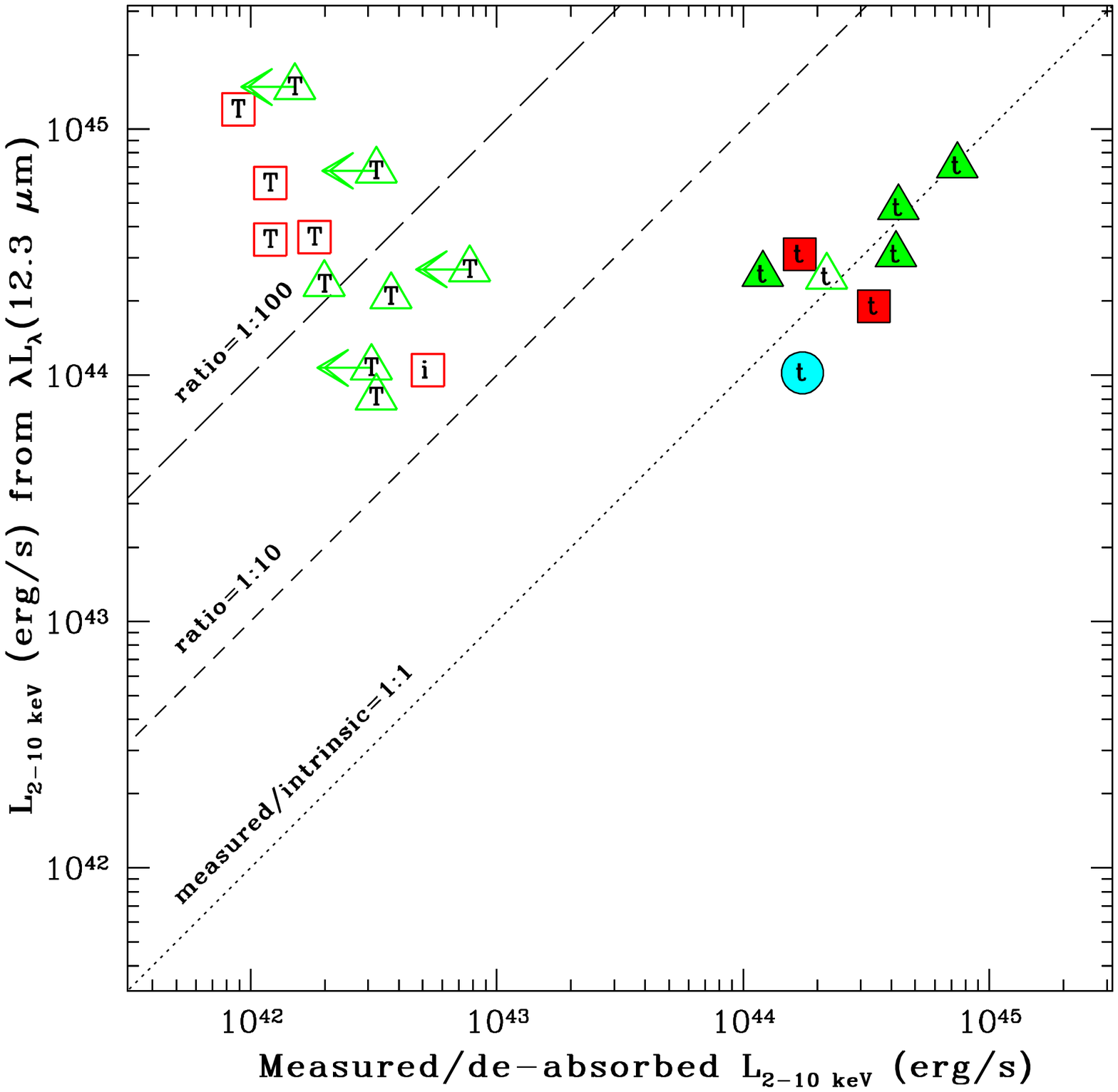}
\caption{
Comparison of the \hbox{2--10~keV} luminosity computed from the 
available \xray\ data with that predicted from the mid-IR luminosity 
at 5.8\micron\ (rest frame) assuming the Fiore et al. (2009) 
correlation {\em (left panel)} and at 12.3\micron\ (rest frame), using 
the Gandhi et al. (2009) correlation {\em (right panel)}.  The dotted, 
short-dashed and long-dashed lines indicate ratios of 1:1, 1:10 and 
1:100 between the measured and the predicted \xray\ luminosity. 
Symbols are the same as in Fig.~\ref{complum}; the filled circle marks 
the source with \rosat\ (RASS) detection (V04).  The letter inside 
each symbol indicates whether the source is a Compton-thick ({\sf T}) 
or a Compton-thin ({\sf t}) AGN candidate on the basis of the 
\oiii-driven results presented in $\S$\ref{discussion} and shown in 
Fig.~\ref{complum}; for one object, the analysis does not lean towards 
any of the two classifications [hence we have an ``intermediate'' 
({\sf i}) AGN], although HR analysis suggests Compton-thin absorption.}
\label{lmir_lx}
\end{figure*}
%%%%%%%%%%%%%%%%%%%%%%%
%%%	End of Figure 3
%%%%%%%%%%%%%%%%%%%%%%%%%%%%%%%%%%%%%%%%%%%%%%%%%%%%%%%%%%%%%%%%%%%%%%%%%%%%%%%%%%%%%

\subsection{Mid-IR luminosity as a proxy of the X-ray emission}
\label{mir_corr}
The rest-frame 5.8\micron\ and 12.3\micron\ luminosities (reported in 
Table~\ref{spitzer}) have been used to estimate the intrinsic, 
rest-frame \hbox{2--10~keV} luminosities ($L_{\rm \scriptsize X,mid-IR}$) 
assuming the relations provided by F09 and 
Gandhi et al. (2009; hereafter G09), respectively. 

The former relation,\footnote{This relation is valid if the criterion 
log \l58$>$43.04 is satisfied, as in the case of all of our sources; 
see F09 for details.} parameterized by $\log L_{\rm 
2-10~keV}$ = 43.57$+$0.72$\times$(log\l58$-$44.2), has been calibrated 
using Type~1 AGN in the \chandra\ Deep Field-South (CDF-S; Brusa et 
al. 2010) and in the \chandra\ Cosmic Evolution Survey 
(\hbox{C-COSMOS}; Civano et al., in preparation); 
for further details, see $\S$3.3 and Fig.~5 in F09. 
This relation is similar (both in slope and intercept) to that 
recently derived by Lanzuisi et al. (2009) from a sample of SWIRE 
sources with \xray\ coverage. For comparison, the assumption of the Lutz et 
al. (2004) relation between the 6\micron\ and hard \xray\ luminosity 
derived using a sample of 71 local, moderate-luminosity Seyfert 
galaxies with low-resolution \iso\ spectra would predict larger \xray\ 
luminosities given the mid-IR luminosities of our sample 
(see below for details).

The latter relation, parameterized by $\log L_{\rm 2-10~keV}$ = 
(log\lmir2$+$4.37)/1.11, has been obtained from a sample of 42 Seyfert 
galaxies with near diffraction-limited mid-IR imaging obtained with the VISIR 
instrument at the {\it VLT} (G09; see also Horst et 
al. 2006, 2008 and 2009). 

All of these correlations are based on the 
assumption that the mid-IR band provides a clear view of the AGN 
nuclear activity, which is more affected by extinction and absorption 
at optical, ultraviolet and \xray\ frequencies. 
However, a few issues must be taken into account in the following discussion. 
\xray\ information is typically obtained from all the available data  
in literature work or in archives 
(from \asca\ to \sax, up to the current \xray\ missions); we note, 
however, that F09 and Lanzuisi et al. (2009) use only 
data with moderate (a few arcsec; \xmm) to high (below 1\arcsec; 
\chandra) resolution, providing overall a relatively good match with 
\spitzer\ imaging data in terms of angular resolution.\footnote{We 
note that source extraction regions of a few tens arcsec are typically 
used in \xmm\ observations.}  On the one hand, G09 
mid-IR data are characterized by sub-arcsec resolution (corresponding 
to physical scales of less than half parsec) matched to a broad 
variety of \xray\ data, including old (\asca) and recent (\suzaku) 
observations.  While host galaxy contribution to hard \xray\ emission, 
at the AGN luminosities probed by most of their sources (above 
10$^{42}$~\lum), is likely negligible (although some reprocessed 
emission may be present), it can still contaminate the mid-IR data at 
some level.  On the other hand, at the median redshift ($z\approx1.5$) 
of the Type~1 AGN used by F09, the mid-IR emission 
would probe physical scales of $\approx$~20~kpc. 
Finally, Lutz et al. (2004) \iso\ spectra are characterized by 
24\arcsec$\times$24\arcsec\ apertures (corresponding to physical 
scales of a few kpc), including, in many cases, noticeable host galaxy 
emission, as pointed out by the authors. 

The results are shown in Fig.~\ref{lmir_lx}, where the 2--10~keV 
luminosity, predicted using the luminosity at 5.8\micron\ (left panel) 
and 12.3\micron\ (right panel) is plotted against the measured \xray\ 
luminosity; 
with the term ``measured \xray\ luminosity'' (similarly to 
Fig.~\ref{complum}), we mean that the source luminosity has been 
corrected for the effect of absorption, once \xray\ spectral analysis 
is able to provide a reliable estimate for it (filled squares: 
this work; filled triangles: V06); in all of the remaining cases, 
no correction has been applied (open squares and triangles). 
In addition, the filled circle represents source ID 239, which was 
detected by the \rosat\ All Sky Survey (RASS; V04).  The dotted, 
short-dashed and long-dashed lines indicate ratios of 1:1, 1:10 and 
1:100 between the measured and the predicted \xray\ luminosity. 
The average \xray\ luminosities predicted by the two methods are 
log~L$_{\rm X}$=44.02~\lum\ and 44.48~\lum\ using the F09 and G09 
mid-IR vs. \xray\ correlation, 
respectively.  For comparison, the average \xray\ luminosity assuming 
the Lutz et al. (2004) relation is log~L$_{\rm X}$=44.41~\lum, 
15~per~cent lower than that predicted using G09 correlation.

Overall, it appears evident that in both plots two main source groups 
are present: the right-most one comprises sources which are more or 
less consistent with the 1:1 correlation, especially using the \lmir2\ as 
a proxy of the nuclear emission (see below for an extended discussion). 
\xray\ spectral analysis indicates that these sources are Compton-thin AGN 
and are the same residing close to the 1:1 correlation  
in Fig.~\ref{complum}. The left-most group regards sources where the 
expected 2--10~keV luminosity is a factor 
\hbox{$\approx$~10--100} higher than the measured \xray\ luminosity. 
All but one of these sources are also among the most extreme objects 
(i.e., candidate Compton-thick AGN) shown in Fig.~\ref{complum}; the 
only exception is source ID 30, which seems to be Compton thin from 
hardness-ratio analysis (see $\S$\ref{spectral_analysis}). 
However, it is evident from Fig.~\ref{lmir_lx} (left panel) that, for the 
Compton-thin Type~2 quasars of our sample, using the correlation 
proposed by F09 provides an offset by a factor $\approx$~3 with 
respect to the \xray\ measured/intrinsic 1:1 luminosity ratio, while 
the G09 correlation seems to produce a good match with the luminosity 
ratio of unity (Fig.~\ref{lmir_lx}, right panel), hence with the 
results obtained from \oiii\ analysis. Although 
clearly mid-IR vs. \xray\ correlations need to be investigated further 
on larger samples to achieve a more accurate ``calibration'', we note 
that, at the measured \xray\ luminosities sampled by the SDSS 
Compton-thin Type~2 quasars (i.e., above 10$^{44}$~\lum), the 
dispersion in the F09 correlation seems to be large (see their 
Fig.~5). By contrast, the correlation presented by G09 has a relative 
small dispersion (0.36 dex) throughout the entire luminosity range, 
being smaller for the subsample of 22 well resolved sources (0.23 dex; 
see $\S$4.1 in G09 for details). Therefore, in the following section, 
we will report on the results obtained using the information from the 
\oiii\ and 12.3\micron\ to predict 2--10~keV luminosities. 

Caution is obviously needed when dealing with infrared data, given the likely
presence of both AGN (torus) and starburst emission.  Although the AGN
seems to dominate over the \spitzer\ bands (eight of the 20 sources
listed in Table~\ref{spitzer} have IRS coverage; see $\S$3.4 of
Zakamska et al. 2008 for details), we note that in the derived
rest-frame 5.8\micron\ and 12.3\micron\ luminosities there might be
some contribution from star formation (see Table~1 of Zakamska et
al. 2008).  However, 
we remind that the correlation reported by G09 was obtained using 
near diffraction-limited mid-IR imaging, where any 
non-AGN contribution can be observed and kept under control (see also 
Horst et al. 2009).

Overall, both \oiii\ and 12.3\micron\ luminosity provide similar 
indications for what concerns the \xray\ emission of SDSS Type~2 
quasars, with roughly half of the sampled (i.e., with \xray\ coverage) 
population being possibly Compton-thick (see $\S$\ref{number_density} 
for further comparison and discussion).  
Recently, heavy obscuration has also been found in some SDSS Type~2 AGN 
by Lamastra et al. (2009) using \xmm\ observations; given the low 
redshifts of their sources, they were able to correct for the extinction 
within the NLR and provide a tight and robust \oiii\ vs. hard 
\xray\ correlation.

Finally, we note that \spitzer\ (IRS) follow-up observations of SDSS 
Type~2 quasars have recently revealed a wealth of mid-infrared 
properties, with 7/12 sources showing a silicate absorption features 
at 10\micron, suggestive of heavy obscuration toward the source 
(Zakamska et al. 2008).  It is intriguing that the two objects with 
the most prominent silicate absorption features (SDSS~0056$+$0032 and 
SDSS~0815$+$4304; see Fig.~1 of Zakamska et al. 2008) are candidate 
Compton-thick quasars accordingly to all methods adopted in this paper 
to estimate the intrinsic \xray\ luminosity, although not all of the 
SDSS Type~2 quasars which have been classified as Compton-thick 
according to our analyses were found to show Si absorption (see Fig.~1 
of Zakamska et al. 2008). This fact can be explained in clumpy 
torus models (e.g., Nenkova, Ivezi{\'c} \& Elitzur 2002; 
H{\"o}nig, Beckert \& Ohnaka 2006; Nenkova et al. 2008a,b), 
where the effective infrared optical depth can be 
decreased by random mis-alignments and holes through the obscuring clouds. 

%%%%%%%%%%%%%%%%%%%%%%%%%%%%%%%%%%%%%%%%%%%%%%%%%%%%%%%%%%%%%%%%%%%%%%%%%%%
%%%	Figure 4: Number density of SDSS Compton-thick QSO candidates 
%%%               vs. other surveys
%%%%%%%%%%%%%%%%%%
\begin{figure}
\includegraphics[angle=0,width=85mm]{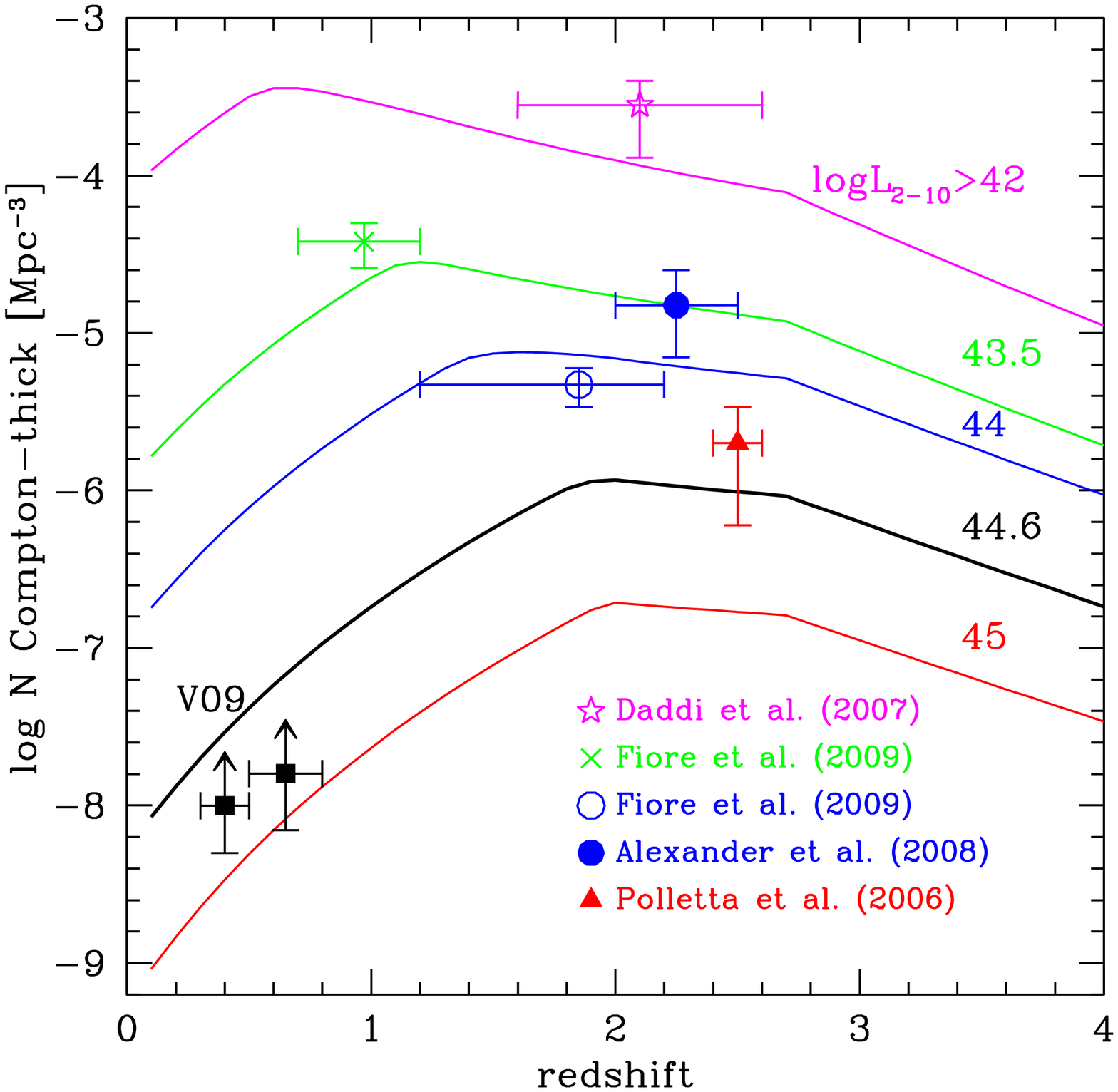}
\caption{
Volume density of Compton-thick AGN samples with different intrinsic 
2--10~keV luminosities compared with the XRB model predictions (Gilli 
et al. 2007). Data points refer to this work (filled squares at low 
redshift with upward-pointing arrows) and literature work (all the 
other symbols, at $z\approx1-2.5$; reference papers are listed in the 
bottom part of the figure), and have to be compared with the 
corresponding color curves.}
\label{num_density}
\end{figure}
%%%%%%%%%%%%%%%%%%
%%%	END of FIG. 4
%%%%%%%%%%%%%%%%%%%%%%%%%%%%%%%%%%%%%%%%%%%%%%%%%%%%%%%%%%%%%%%%%%%%%%%%%%%

%%%%%%%%%%%%%%%%%%%%%%%%
%%%%%%%%%%%%%%%%%%%%%%%%%%%%%%%%%%%%%%%%%%%%%%%%%%%%%%%%%%%%%%%%%%%%%%%%%%%%%%%%%%
%%%	Table 5: SDSS Type 2 QSOs used to compute the Compton-thick Number Density
%%%%%%%%%%%%%%%%
\begin{table*}
\centering
\begin{minipage}{145mm}
\caption{Classification of Type~2 quasars with log$L_{\rm [OIII]}>9.28$~\lsun.}
\label{summary_ct}
\begin{tabular}{ccccccc}
\hline
ID         & Src. Name            & $z$   & $L_{\rm \scriptsize X, meas}$/$L_{\rm \scriptsize X, [OIII]}$  
                                                        & $L_{\rm \scriptsize X, meas}$/$L_{\rm X, \scriptsize 12.3\micron}$ 
                                                                      & Class.~\oiii\ 
                                                                                 & Class.~12.3\micron\ \\
(1)        &   (2)                & (3)   &  (4)        &   (5)       &   (6)       &  (7)     \\
\hline
\phn\phn 7 & 005009.81$-$003900.6 & 0.729 & 0.37        & \dots       & thin     &  \dots  \\
\phn\phn 9 & 005621.72$+$003235.8 & 0.484 & 1.41E-03    & 7.41E-04    & THICK    &  THICK     \\
\phn    16 & 012032.21$-$005502.0 & 0.601 & 4.90E-03    & \dots       & THICK    &  \dots  \\
\phn    20 & 013416.34$+$001413.6 & 0.555 & 1.58E-03    & 3.39E-03    & THICK    &  THICK     \\
\phn    29 & 014932.53$-$004803.7 & 0.566 & $<$4.27E-03 & \dots       & THICK    &  \dots  \\
\phn    30 & 015716.92$-$005304.8 & 0.422 & 7.08E-03    & 5.01E-02    & interm   &  thin      \\
\phn    34 & 021047.01$-$100152.9 & 0.540 & 0.16        & 0.87        & thin     &  thin      \\
\phn    83 & 031950.54$-$005850.6 & 0.626 & 4.37E-03    & 1.78E-02    & THICK    &  thin     \\
       100 & 073745.88$+$402146.5 & 0.613 & 7.08E-03    & \dots       & THICK    &  \dots  \\
       113 & 080154.24$+$441234.0 & 0.556 & 0.49        & 1.3         & thin     &  thin      \\
       119 & 081507.42$+$430427.2 & 0.510 & $<$3.89E-03 & $<$4.79E-03 & THICK    & THICK      \\
       130 & 084234.94$+$362503.1 & 0.561 & $<$2.75E-03 & $<$2.88E-02 & THICK    & interm     \\
       153 & 092152.45$+$515348.1 & 0.587 & $<$5.37E-03 & \dots       & THICK    &  \dots  \\
       182 & 102746.03$+$003205.0 & 0.614 & 3.09E-03    & \dots       & THICK    &  \dots  \\
       186 & 103951.49$+$643004.2 & 0.402 & 2.09E-03    & 2.00E-03    & THICK    &  THICK     \\
       196 & 115314.36$+$032658.6 & 0.575 & 0.13        & 0.47        & thin     &  thin      \\
       197 & 115718.35$+$600345.6 & 0.491 & $<$1.70E-03 & $<$1.02E-03 & THICK    &  THICK     \\
       204 & 122656.48$+$013124.3 & 0.732 & 0.42        & 0.89        & thin     &  thin      \\
       205 & 122845.74$+$005018.7 & 0.575 & 8.13E-02    & \dots       & thin     &  \dots  \\
       207 & 123215.81$+$020610.0 & 0.480 & 1.82E-03    & 8.51E-03    & THICK    &  THICK     \\
       232 & 144642.29$+$011303.0 & 0.725 & 9.55E-03    & \dots       & THICK    &  \dots  \\
       244 & 151711.47$+$033100.2 & 0.613 & 2.95E-03    & \dots       & THICK    &  \dots  \\
       256 & 164131.73$+$385840.9 & 0.596 & 0.40        & 1.0         & thin     &  thin      \\
       289 & 235818.87$-$000919.5 & 0.402 & $<$6.61E-03 & $<$2.88E-02 & THICK    &  interm    \\
       290 & 235831.16$-$002226.5 & 0.628 & 1.58E-03    & 3.98E-02    & THICK    &  thin      \\
\hline
\end{tabular}
\end{minipage}
\hglue-1.0cm
\begin{minipage}{130mm}
(1) Source ID (from Table~1 of Z03); 
(2) SDSS source name; 
(3) source redshift; 
(4) ratio of the measured 2--10~keV luminosity and that predicted from the \oiii5007\AA\ luminosity assuming the M94 correlation; 
(5) ratio of the measured 2--10~keV luminosity and that predicted from the 12.3\micron\ luminosity (when available; see Table~\ref{spitzer}) 
    assuming the G09 correlation; 
(6) tentative source classification on the basis of the measured/predicted (from the \oiii) luminosity ratio. 
    For source \#30, although a Compton-thick classification is derived from the luminosity ratio, 
    hardness-ratio analysis ($\S$\ref{spectral_analysis}) suggests Compton-thin obscuration; 
(7) tentative source classification on the basis of the measured/predicted (from the 12.3\micron) luminosity ratio; 
    objects are classified as ``intermediate'' if the upper limit on the luminosity ratio prevents any conclusion. \\
%A ``?'' close to the classification indicates 
%that the corresponding luminosity ratio is very close to the chosen arbitrary boundary of Compton-thick vs. Compton-thin quasar. 
\end{minipage}
\end{table*}
%%%%%%%%%%%%%%%%
%%%    End of Table 5
%%%%%%%%%%%%%%%%%%%%%%%%%%%%%%%%%%%%%%%%%%%%%%%%%%%%%%%%%%%%%%%%%%%%%%%%%%%%%%%%%%%%%%%%%%

%%%%%%%%%%%%%%%%%%%%%%%%

\section{Number density of SDSS Compton-thick quasars}
\label{number_density}
It is now interesting to compute the space density of candidate 
Compton-thick Type~2 quasars selected by the SDSS and compare it with 
that of other samples of Compton-thick AGN candidates and with the 
predictions by synthesis models of the XRB.  First, we considered the 
complete sub-sample of radio-quiet objects presented in V06 and in 
this work with $\log$\loiii$>9.28$~\lsun\ 
(therefore source SDSS~0812$+$4018 has been removed because it is 
radio loud; see $\S$\ref{chandra_obs}): 
six objects fall in the 
redshift range $z=0.3-0.5$ [with 5 objects being Compton-thick 
candidates using the \xray\ measured/predicted (from \oiii) luminosity 
ratio, $L_{\rm \scriptsize X, meas}$/$L_{\rm \scriptsize X, [OIII]}$] 
and 19 objects fall in the redshift range $z=0.5-0.8$ (with 12 
objects being Compton-thick candidates, still according to \oiii); the 
properties of these sources are summarized in Table~\ref{summary_ct}.

Before entering into the details of the number density 
computation, we remind the reader that we refer to a Compton-thick 
quasar when the ratio of the measured to predicted \xray\ luminosity 
is less than 0.01.  On the basis of the $L_{\rm \scriptsize X, 
meas}$/$L_{\rm \scriptsize X, [OIII]}$ ratio, the \xray\ emission for 
17 of the 25 Type~2 quasars in our sample (68~per~cent) 
appears to be obscured by Compton-thick material. However, since the 
Type~2 quasar sample is selected on \oiii\ luminosity, the estimated 
Compton-thick quasar fraction may be overestimated.  We can provide an 
indication of the ``\oiii\ bias'' by determining the fraction of the 
Type~2 quasars that are classified as Compton thick based on 
both the \xray/\oiii\ and \xray/mid-IR ratios. Of the 16 Type~2 
quasars with archival \spitzer\ observations and 
$\log$\loiii$>9.28$~\lsun\ (see Table~\ref{summary_ct}), 10 have 
$L_{\rm \scriptsize X, meas}$/$L_{\rm \scriptsize X, [OIII]}$$<$0.01, 
of which 6 (60~per~cent) also have $L_{\rm \scriptsize X, 
meas}$/$L_{\rm \scriptsize X, mid-IR}$$<$0.01 (where mid-IR is based 
on 12.3\micron); however, two objects have 
\xray/mid-IR upper limits which are slightly above our Compton-thick 
quasar definition (``intermediate'' in Table~\ref{summary_ct}), 
suggesting that the fraction of Compton-thick 
quasars may be 75~per~cent (i.e., 6 out of 8 objects). 

Overall, our analyses therefore suggest that the unbiased 
Compton-thick quasar fraction is $\approx$~50~per~cent.\footnote{The 
current study indicates that, for the 14 sources with \oiii\ and 
12.3\micron\ indicators (16 objects minus two classified as ``intermediate'' 
in the mid-IR), \oiii\ is highly reliable in classifying sources as 
Compton-thick in at least 75~per~cent of the cases (since source \#83 
has a ``borderline'' classification as Compton thin according to the 
12.3\micron\ indicator).  Applying this ``correction factor'' to the 
number of Compton-thick sources with only \oiii\ as reliable indicator 
(i.e., no solid mid-IR information) implies at least 7 Compton-thick 
quasars out of 9 (hence, 13/25 for the sample under investigation).}

In our calculation below, we will refer to the \oiii\ indicator to 
compute the number density of SDSS Compton-thick quasars, since it is 
available for all of the 25 sources under consideration; however, any 
correction to the \oiii\ bias would decrease our estimate by 
$\approx$~25~per~cent at most. 

Recently, R08 published an extended list of SDSS Type~2 AGN candidates. 
They also reported new values for the \oiii\ line luminosity, 
which is measured using both a fitting with a Gaussian feature and with a 
non-parametric method (see Section~2.4 of R08 for a detailed description), 
besides adopting the most up-to-date spectro-photometric calibration algorithm 
from the SDSS. 
Since our project was originally intended to provide an \xray\ 
characterization of all of the sources of the Z03 sample 
in the quasar locus (i.e., at high \xray\ luminosities, 
predicted accordingly to the \oiii\ vs. hard \xray\ correlation of M94; 
see Fig.~\ref{lxoiiiz}), in the following 
we will keep the original \oiii\ line luminosity reported by Z03. 
However, it is interesting to analyze thoroughly what the main differences 
would be in case of source selection on the basis of the \oiii\ line 
luminosity as reported in R08. 
If we adopt the same selection of $\log$\loiii$>9.28$~\lsun\ using the R08 
source catalog instead of the original catalog of Z03, 
we obtain a sample comprising 26 sources, 22 in common with the source list 
reported in Table~\ref{summary_ct}. On the one hand, sources \#16, \#30 and 
\#244 (two classified as Compton thick on the basis of \oiii\ only, and one as 
intermediate/Compton-thin) would not be present anymore for our estimates 
of the volume density of SDSS Compton-thick quasars. 
On the other hand, sources \#117, \#141, \#189 and \#213 would be included. 
While source \#117 must be removed because of its radio loudness 
(its exclusion was described at the beginning of this Section), 
a similar destiny would be reserved to source \#189, for the same reason. 
The other two sources lack any \xray\ constrain to date, so they would not 
be useful for the following analysis. 
In the light of the limited number of changes implied by the adoption 
of the R08 catalog (and the way we account for the Compton-thick AGN 
predicted using only \oiii, as described above), 
we are confident that the main conclusions of the present work 
(reported in the following paragraphs) will not change significantly. 
We also note that, for the sources above the chosen \oiii\ line 
luminosity threshold in both catalogs (22 objects), the 
logarithms of the luminosities derived for \oiii\ differ, on average, 
by $\approx$~0.09, with no obvious evidence for systematic effects. 

In order to get the space density of Type~2 quasars above the chosen 
luminosity threshold, we integrated the SDSS Type~2 quasars luminosity 
function published by R08 from 
$\log$\loiii$>9.28$~\lsun\ to the maximum luminosity for which they 
have data, i.e. $\log$\loiii=10.3~\lsun\ and $\log$\loiii=10.0~\lsun\ 
for objects in the range $z=0.3-0.5$ and $z=0.5-0.8$, respectively. 
For the objects in the $z=0.5-0.8$ bin, we considered both curves in 
Fig.~5 of R08, in which the measured space densities depend on a 
different correction applied to the Type~2 quasar selection function 
(see R08 for details). These integrations return a space density of 
$1.2\times10^{-8}$~Mpc$^{-3}$ for Type~2 quasars at $z=0.3-0.5$ and 
$(1.9-2.6) \times 10^{-8}$~Mpc$^{-3}$ (depending on the correction 
mentioned above) for Type~2 quasars at \hbox{$z=0.5-0.8$}. Since 5 out 
of 6 objects in our $z=0.3-0.5$ sample and 12 out of 19 objects in our 
$z=0.5-0.8$ sample are Compton-thick candidates, we estimate the space 
densities of \oiii-selected SDSS Compton-thick quasars to be 
$1.2\times 10^{-8}\times (5/6) = 10^{-8}$~Mpc$^{-3}$ and 
$(1.9-2.6)\times 10^{-8}\times (12/19) = (1.2-1.6) \times 
10^{-8}$~Mpc$^{-3}$, at $z=0.3-0.5$ and $z=0.5-0.8$, 
respectively. Based on the observed \oiii\ vs. intrinsic $L_{2-10 
keV}$ relations published by M94, the above space densities should 
then refer to Type~2 quasars with $logL_X>44.6$~\lum. 
In Fig.~\ref{num_density} we show the comparison between the space 
densities derived for our sample and those published for other samples 
of Compton-thick AGN candidates at different redshifts and 
luminosities. The densities presented in this work cover a corner in 
the luminosity vs. redshift plane of Compton-thick AGN so far 
unexplored. We also show the expected redshift evolution of the space 
density of Compton-thick AGN with different intrinsic luminosities 
based on the XRB model of Gilli et al. (2007). Our measurements are 
about a factor of 2--3 lower than the expectations of the XRB model, in 
which Type~2 Compton-thick quasars are assumed to be as abundant as 
unobscured quasars with the same intrinsic luminosity. 
A number of Compton-thick AGN smaller than expected 
(by a factor of 2, corresponding to $\approx$~2$\sigma$) 
might also be suggested by recent \integral\ (Beckmann et al. 2009) 
and \swift\ (Tueller et al. 2008) surveys of very local objects, 
which would point towards a revision of the XRB models 
(although this issue is still matter of debate, see Malizia et al. 2009). 
For instance, a lower space density of Compton-thick AGN features the 
recent model by Treister, Urry \& Virani (2009). 

A few caveats have nonetheless to be 
kept in mind: indeed, R08 consider their estimates to be lower limits, 
since, if the NLR is extincted, a significant fraction of obscured AGN 
may escape detection based on narrow optical emission lines (other 
arguments are discussed in R08).  In addition, the intrinsic \xray\ 
luminosity of these Compton-thick Type~2 quasar candidates has been 
obtained using an observed \oiii\ vs. intrinsic \xray\ luminosity 
relation derived by M94 for a local sample of Seyferts, 
which generally cover a lower \oiii\ luminosity range, 
and have not been selected on the basis of their 
\oiii\ flux or luminosity. 
As mentioned in $\S$\ref{discussion}, other relations 
between \oiii\ and \xray\ luminosities have been published by H05 and 
P06, but none of them is based on a purely \oiii\ selected sample, 
although H05 made an attempt to build a sample of \oiii-bright local 
Seyferts as a surrogate.  For their sample of Seyfert~1 objects, 
i.e. for those sources in which the intrinsic \xray\ emission can be 
estimated directly, H05 derived \xray\ luminosities on average a 
factor of 1.5 lower than those derived by M94 at a given \oiii\ 
luminosity [$log\;L_{X}/L_{\rm [OIII]}$=1.59 vs 1.76]. Therefore, 
using the relation by H05 we would estimate for our sources intrinsic 
luminosities of $logL_X>44.4$~\lum.  In the considered XRB synthesis 
model, the space density of obscured AGN, and therefore also of 
Compton-thick AGN, is expected to increase with decreasing intrinsic 
luminosity. At $z=0.3-0.8$, the XRB space density of Compton-thick AGN 
with $logL_X>44.4$~\lum\ is expected to be about an order of magnitude 
larger than that measured in this work.  On the other hand, when using 
the relation by P06, which is essentially based on optically selected 
AGN samples, the \oiii\ luminosity threshold translates into an \xray\ 
luminosity threshold of $logL_X>44.8$~\lum. At these luminosities, the 
model space density of Compton-thick Type~2 quasars is very close to 
our measurements. Summarizing, given the uncertainties mentioned above 
(see also $\S$~\ref{discussion}), at $z=0.3-0.8$, Compton-thick Type~2 
quasars selected by the SDSS are compatible with being as abundant as 
unobscured quasars of similar luminosities (see also Alexander et al. 2008).

\section{Summary}
\label{summary}
We have presented the most up-to-date results regarding the properties 
of Type~2 quasar candidates selected from the SDSS and observed by 
\chandra. In particular, the objects presented in this work consists 
on two samples, one comprising 12 Type~2 quasars selected in \chandra\ 
Cycle~8 among the most luminous \oiii\ emitters, and the other 
comprising three additional candidates at lower \oiii\ luminosity 
retrieved from the \chandra\ archive. The full sample spans the 
redshift range \hbox{$z$=0.40--0.73}.  The main results are summarized below: 
%%%%%%%%%%%%%%%%%%%
\begin{description}
%%%%%%%%%%%%%%%%%%%
\item$\bullet$ 
Our \chandra\ exploratory observations ($\approx$~10~ks for each 
pointing) support previous findings that SDSS Type~2 quasars are 
\xray\ faint: two sources were not detected by \chandra\ and nine of 
the detected sources have less than 10 counts in the observed 
\hbox{0.5--8~keV} band.  For only three sources it was possible to 
perform a low-to-moderate quality spectral analysis, providing 
evidence for column densities in the range 
$N_{\rm H}\approx$~10$^{22}$--10$^{23}$~cm$^{-2}$ in the source rest frame. 
%%%%%%%%%%%%%%
\item$\bullet$ 
%%%%%%%%%%%%%%
The assumption of the \oiii\ luminosity as a proxy of the nuclear 
emission, coupled with literature \oiii--hard 
\xray\ correlations (with all the caveats of this approach being 
extensively discussed in $\S$\ref{discussion} 
and $\S$\ref{number_density}), indicates that a significant fraction 
(about two-third) of the SDSS Type~2 quasars with 
$\log$\loiii$>9.28$~\lsun\ presented here and in V06 may be Compton 
thick (see also Ptak et al. 2006 and Lamastra et al. 2009).  
%%%%%%%%%%%%%%
\item$\bullet$ 
%%%%%%%%%%%%%%
Further indications of heavy obscuration in a sizable fraction of SDSS 
Type~2 quasars come from the analysis of archival \spitzer\ 
observations, which have been used to estimate the intrinsic (i.e., 
de-absorbed) hard \xray\ AGN strength assuming some recent 
correlations between either the rest-frame 5.8\micron\ or the 
12.3\micron\ luminosity and the 2--10~keV luminosity. Using the 
combined \oiii, mid-IR and \xray\ information (by means of luminosity 
ratios; see Table~\ref{summary_ct}) it is possible to distinguish the 
Compton-thick from the Compton-thin Type~2 quasar population. 
Overall, we find that $\approx$~50~per~cent of SDSS Type~2 quasars 
with $\log$\loiii$>9.28$~\lsun\ appear to be obscured by Compton-thick 
material based on both the $L_{\rm \scriptsize X, meas}$/$L_{\rm 
\scriptsize X, mid-IR}$ (where mid-IR corresponds to rest-frame 12.3\micron) 
and $L_{\rm \scriptsize X, meas}$/$L_{\rm \scriptsize X, [OIII]}$ ratios. 
Moderate-quality \xray\ spectra showing a strong iron K$\alpha$ 
emission line over a flat continuum would provide a further, clean 
indication of Compton-thick absorption. However, given the low 
\xray\ fluxes of these sources, such spectral approach, extended to  
sizable samples, probably demands next-generation of sensitive \xray\ 
detectors.
%%%%%%%%%%%%%%
\item$\bullet$ 
%%%%%%%%%%%%%%
By integrating the SDSS Type~2 quasar luminosity function of R08 above 
log$L_{\rm [OIII]}>9.28$~\lsun\ (corresponding to $logL_X>44.6$~\lum\ 
under the assumption of the M94 relation; see Fig.~\ref{lxoiiiz}), 
%in two redshift intervals ($z=0.3-0.5$ and $z=0.5-0.8$), 
we obtain an estimate of the space density of Type~2 quasars above 
this luminosity threshold of $\approx10^{-8}$~Mpc$^{-3}$ and 
$\approx(1.2-1.6)\times 10^{-8}$~Mpc$^{-3}$ at $z=0.3-0.5$ and $z=0.5-0.8$, 
respectively. 
At face value, these space densities are a factor 2--3 lower than the 
expectations of the XRB model by Gilli et al. (2007) and would suggest 
that Compton-thick Type~2 quasars are less abundant than unobscured 
quasars (Treister et al. 2009). However, given the 
uncertainties in the SDSS Type~2 quasar luminosity function (mainly 
related to possible extinction within the NLR) and in the assumption 
of the \oiii-hard \xray\ correlation (see $\S$\ref{number_density}), 
we can conclude that the estimated space density of SDSS Compton-thick 
Type~2 quasars at $z=0.3-0.8$ is likely a lower limit. 
In other words, in this still poorly explored redshift range (recent 
results on space densities of heavily obscured quasars are mostly at 
$z\approx1-2.5$), Compton-thick Type~2 quasars are consistent with 
being as abundant as unobscured quasars of similar intrinsic luminosities. 
\end{description}
%%%%%%%%%%%%%%%%%%%%

\section*{Acknowledgments}
CV and RG thank for partial support the Italian Space Agency 
(contracts ASI--INAF I/023/05/0 and ASI I/088/06/0) and PRIN--MIUR 
(grant 2006-02-5203). DMA acknowledges support by the Royal Society 
and the Leverhulme Trust. The authors thank S. Bianchi, A. Comastri, 
R. Della Ceca, P. Gandhi, A. Lamastra, G. Matt, F. Panessa, 
G.~C. Perola, M. Polletta and G. Zamorani for useful suggestions, 
and M. Mignoli for help with SDSS Type~2 quasar spectra, and the anonymous 
referee for his/her careful reading of the manuscript and useful comments.

\end{document}